\documentclass[fleqn,usenatbib]{mnras}

\usepackage{newtxtext,newtxmath}
\usepackage[T1]{fontenc}
\usepackage{tabularx}

\newcommand\clearrow{\global\let\rowmac\relax}
\clearrow
\usepackage{natbib}
\usepackage{graphicx}	% Including figure files
\usepackage{amsmath}	% Advanced maths commands
\defcitealias{vdMA10}{vdMA10}
\defcitealias{NGB08}{NGB08}
\defcitealias{N10}{N10}

\title[$\omega$~Centauri MUSE analysis]{$\omega$~Centauri: A MUSE discovery of a counter-rotating core\thanks{Based on observations collected at the European Organisation for Astronomical Research in the Southern Hemisphere under ESO programs 0103.D-0204, 0104.D-0257, 105.20CR, and 109.23DV}}

\author[Pechetti et al.]{Renuka Pechetti,$^{1}$
Sebastian Kamann,$^{1}$ 
Davor Krajnovi\'c,$^{2}$ 
Anil Seth,$^{3}$ 
Glenn van de Ven,$^{4}$
\newauthor
Nadine Neumayer,$^{5}$ 
Stefan Dreizler,$^{6}$ 
Peter M. Weilbacher,$^{2}$ 
Sven Martens,$^{6}$
Florence Wragg$^{1}$
\\
$^{1}$Liverpool John Moores University, UK\\
$^{2}$Leibniz--Institut f\"ur Astrophysik Potsdam (AIP)\\
$^{3}$University of Utah, Salt Lake City, Utah\\
$^{4}$University of Vienna, Vienna, Austria\\
$^{5}$Max Planck Institut f\"ur Astronomie, Heidelberg, Germany\\
$^{6}$Institut f\"ur Astrophysik, G\"ottingen, Germany
}

\pubyear{2024}

\begin{document}

\label{firstpage}
\pagerange{\pageref{firstpage}--\pageref{lastpage}}
\maketitle

\begin{abstract}
$\omega$~Centauri is considered the most massive globular cluster of the Milky Way and likely the former nuclear star cluster of a galaxy accreted by the Milky Way. It is speculated to contain an intermediate-mass black hole (IMBH) from several dynamical models. However, uncertainties regarding the location of the cluster center or the retention of stellar remnants limit the robustness of the IMBH detections reported so far. In this paper, we derive and study the stellar kinematics from the highest-resolution spectroscopic data yet, using the Multi Unit Spectroscopic Explorer (MUSE) in the narrow field mode (NFM) and wide field mode (WFM). Our exceptional data near the center reveal for the first time that stars within the inner 20" ($\sim$0.5~pc) counter-rotate relative to the bulk rotation of the cluster. Using this dataset, we measure the rotation and line-of-sight velocity dispersion (LOSVD) profile out to 120$''$ with different centers proposed in the literature. We find that the velocity dispersion profiles using different centers match well with those previously published. Based on the counter--rotation, we determine a kinematic center and look for any signs of an IMBH using the high-velocity stars close to the center. We do not find any significant outliers~$>$60~km/s within the central 20$''$, consistent with no IMBH being present at the center of $\omega$~Centauri. A detailed analysis of Jeans' modeling of the putative IMBH will be presented in the next paper of the series.

\end{abstract}


%%%%%%%%%%%%%%%%%%%%%%%%%%%%%%%%%%%%%%%%%%%%%%%%%%

%%%%%%%%%%%%%%%%% BODY OF PAPER %%%%%%%%%%%%%%%%%%

\section{Introduction}

The most massive cluster of the Milky Way, $\omega$~Centauri (NGC~5139), has been a topic of discussion for more than four decades now \citep[e.g.,][]{freeman75,geyer83,meylan95,lee02}. Many studies have also theorized that it might be the surviving nucleus (or nuclear star cluster, NSC) of a stripped dwarf galaxy due to the presence of complex stellar populations that display a broad metallicity distribution \citep[e.g.,][]{j&p10,husser20}, and are far more complex than the multiple populations routinely found in other clusters \citep[e.g.,][]{gratton12,piotto15,martocchia18}. It also has evidence for a central stellar disk and tangential velocity anisotropy consistent with tidal stripping \citep[e.g.,][]{vandeven06}. More recently, data from the Gaia satellite was used to trace the origin of Galactic globular clusters (GCs)\citep{massari19}, and $\omega$~Centauri was suggested to be the former core of the $Gaia$-Enceladus/Sausage galaxy \citep{forbes20,pfeffer21}, which was a dwarf galaxy with a mass of $\sim$10$^8$~M$_\odot$ accreted by the Milky Way $\sim$10~Gyr ago \citep{helmi18}.

Several stripped nuclei, sometimes known as ultra-compact dwarf galaxies (UCDs), were recently detected around nearby galaxies. These remnants of more massive galaxies~($>$10$^9$~M$_\odot$) are capable of hosting supermassive black holes with masses~$>10^6~$M$_\odot$ \citep[e.g.,][]{seth14,ahn17,ahn18}. Recent analyses have further shown that somewhat less massive black holes (BHs) are present in all five of the nearest early-type galaxies with stellar masses $\sim$10$^9$~M$_\odot$ and NSC masses between 2$\times$10$^6$--7$\times$10$^7$~M$_\odot$ \citep{ngyuen17,nguyen18,nguyen19}. Therefore, one might expect a high fraction of stripped NSCs from this mass range of galaxies to also host massive BHs~$<$10$^6$~M$_\odot$.
%These UCDs represent the highest luminosity end of the globular cluster and stripped nuclei population. The highest These discoveries indicate 
% This makes it likely that the low-mass stripped nuclei are remnants of dwarf galaxies ($<$10$^9$~M$_\odot$), which have a greater chance of hosting intermediate-mass black holes (IMBHs).
 Indeed, a 10$^5$ M$_\odot$ BH was recently found in the M31 globular cluster (GC), B023-G078 \citep{pechetti22}.  Like $\omega$~Centauri, this object has additional evidence of being a stripped nucleus. There have been other proposed detections, for example, G1 in M31 \citep{gebhardt05},  M54 \citep{ibata09}, the likely nuclear star cluster of the Sagittarius dwarf galaxy \citep[e.g.,][]{alfaro-cuello19}. This establishes that GCs are potential sites of IMBHs despite the lack of robust detections in Milky Way GCs.
% Recently, a $>$3$\sigma$ detection of an IMBH in the most massive cluster of M31, B023-G078 \citep{pechetti22} was reported, along with strong evidence for the cluster to be a stripped nucleus.  
Completing our picture of the number density of massive BHs in the cosmic neighborhood is a crucial step towards  understanding the formation of the seed BHs in the early universe \citep[e.g.,][]{volonteri10,greene20} as well as the correlations between galaxies and their BHs \citep[e.g.,][]{saglia16,habouzit21}. 

$\omega$~Centauri is a perfect candidate to search for an IMBH since it is the most massive cluster in the Milky Way, with a dynamical mass of 2.5 -- 3.5$\times$10$^6$~M$_\odot$ \citep{vandeven06,baumgardt18}, and an outlier in the globular cluster luminosity function \citep{kruijssen09}. Like M54, it is an outlier from regular globular clusters in the relation between average metallicity and intrinsic metallicity spread  (e.g., Figure 2 of \citet{leaman12}), where both lie on the dwarf galaxy sequence. It is strongly rotating with a rotation velocity of $\sim$4--5~km/s \citep{mm86,merritt97,sollima19} and has one of the highest central velocity dispersions of $\sim$22~km/s  \citep{N10}, which makes it an outlier in the $V/\sigma$ among the Galactic GCs. The rise in the velocity dispersion in \citep{N10} suggests the presence of an IMBH, but an IMBH is not the only solution. Several analyses, such as \citet{baumgardt19} and \citet{zocchi19} have shown that this rise could also be caused by the presence of an extended mass distribution consisting of stellar-mass BHs instead of a single IMBH. \citet{zocchi19} further show that if radial anisotropy near the center is considered, a central extended dark mass of~$<$5\% of the cluster mass is sufficient to explain the observed kinematics. Other studies have also proposed the possibility of concentrated non-baryonic matter present in the core of $\omega$~Centauri. For example, assuming a Navarro-Frenk-White (NFW) profile, \citet{brown19} find an integrated dark mass of $\sim$5$\times$10$^5$~M$_\odot$ at the cluster center. \citet{evans22} performed an analysis with different dark matter profiles for the central dark mass in the cluster and found that although stellar remnants can explain masses~$<$5$\times$10$^5$~M$_\odot$, any mass greater than that cannot be explained by it. Although various interpretations can be given for the central dark mass component, no conclusions have been found robustly yet. 
 
A major source of uncertainty in dynamical measurements are the central density slope and the velocity dispersion profile, which changes based on the adopted center of the cluster, of which several estimates exist from \citet{NGB08,N10,vdMA10}, which are hereafter referred to as \citetalias{NGB08}, \citetalias{N10} and \citetalias{vdMA10} respectively. These studies estimate centers based on kinematics or photometry and from different datasets such as integrated kinematics and proper motions of the stars. The \citetalias{NGB08} center was determined based on the central density of $\omega$~Centauri and is $\sim$12$''$ away from \citetalias{N10} and \citetalias{vdMA10} center. The \citetalias{N10} and \citetalias{vdMA10} centers are  $\sim$3$''$ apart but were determined using kinematics and number density counts respectively. These centers have produced different dispersion profiles, for example, the profile in \citetalias{N10} peaks strongly resulting in a central velocity dispersion of $\sim$22~km/s whereas the dispersion profile of \citetalias{vdMA10} is relatively flatter at $\sim$~19~km/s towards the center. This has also resulted in different estimates for the IMBH mass where \citetalias{NGB08} and \citetalias{N10} argued for the presence of an IMBH with a mass of $4~-~5\times~10^4\,{\rm M_\odot}$, whereas \citetalias{vdMA10} found that no IMBH was required to fit the observed kinematics of the cluster.  As noted by \citetalias{vdMA10} and \citetalias{N10}, the exact location of the center along with the kinematics based on that center are necessary for arguing the presence of an IMBH.

 A detailed analysis of the kinematics is thus required to solve the discrepancies regarding the center. With $\omega$~Centauri being the most extensively studied cluster in the Milky Way, several kinematic datasets from different instruments exist. We have obtained the highest spatial resolution data (50~mas) yet in the central 20$''$ using integrated field MUSE-NFM spectroscopy. We also obtained MUSE-WFM spectroscopy for $\sim$40,000 stars within the half-light radius of this cluster to quantify the presence of a dark mass, either in the form of an IMBH \citep{N10,baumgardt17} or a collection of stellar mass BHs that have mass segregated to the central regions of the cluster \citep{baumgardt19,zocchi19}. We revisit two crucial aspects in this paper that are required to probe the presence (and constrain the potential mass) of an IMBH in $\omega$~Centauri, namely the central kinematics as well as the determination of the cluster center. The latter will also enable an accurate determination of the surface brightness profile. We analyse the kinematics based on the different existing centers proposed in the literature and attempt to determine the kinematic center based on the detection of a centralized rotation signal. In a subsequent paper, an analysis of the presence/absence of the IMBH is done based on the Jeans' modeling of the kinematics.

% We use high-resolution adaptive optics data obtained with the MUSE integral-field spectrograph \citep{bacon10} to derive the central kinematics of $\omega$~Centauri using the line-of-sight (LOS) velocities of cluster members.  We analyse the kinematics based on the different existing centers proposed in the literature and attempt to determine the kinematic center based on the detection of a centralized rotation signal.}

This paper is organized as follows. In Section~2 we present the imaging and spectroscopic data. Section~3 presents the analysis of the kinematic data based on the different centers and Section~4 describes the analysis of the inner 20$''$ of the cluster. Section~5 consists of a discussion on the existence of high-velocity stars in $\omega$~Centauri. Section~6 contains our summary and conclusions.

\begin{figure*}
    \includegraphics[trim={0.2cm 0cm 0.2cm 0.8cm},clip,width=\linewidth]{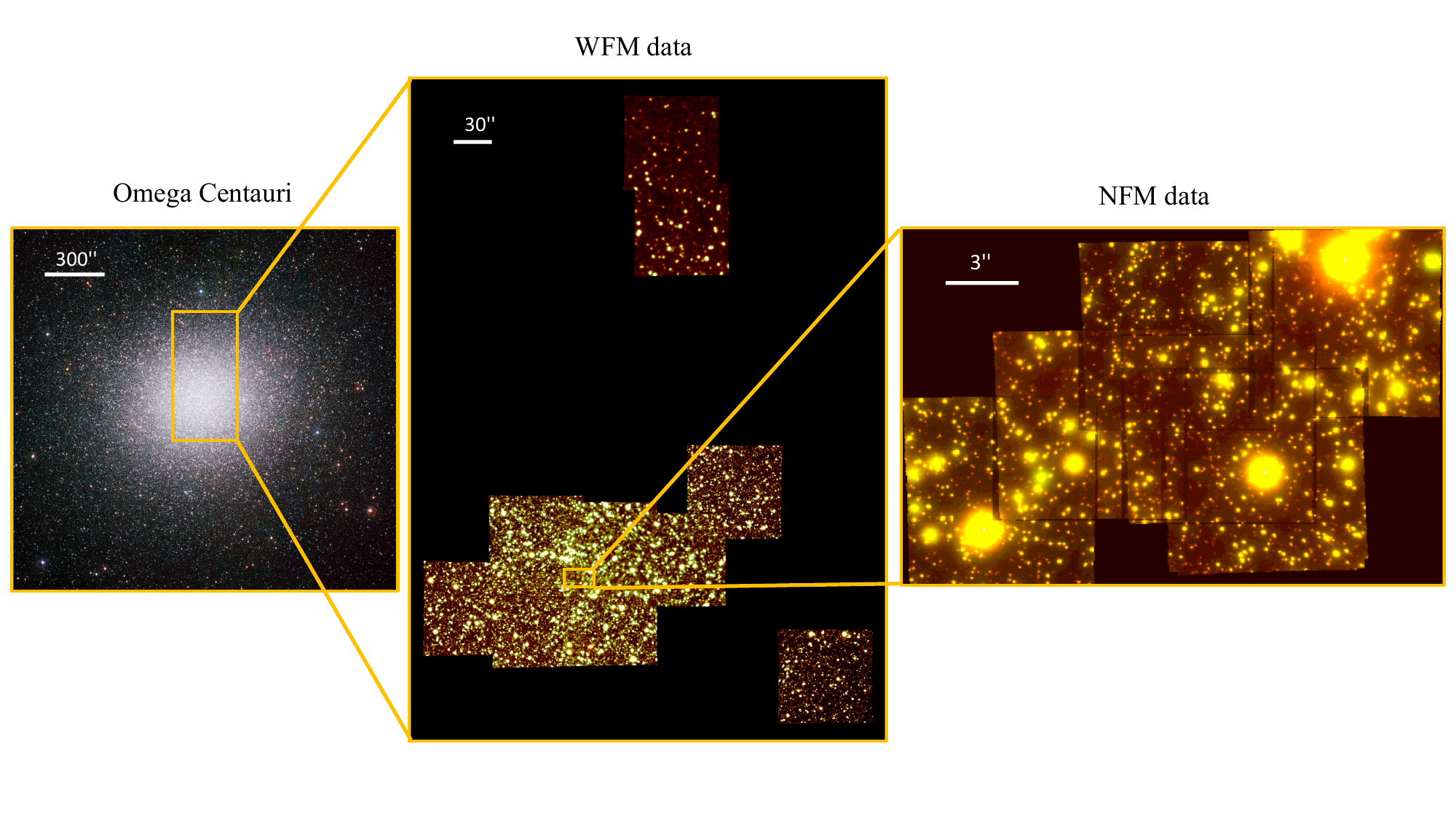}
     \caption{MUSE data of $\omega$~Centauri. The left panel shows a  0.5$^\circ$ image of $\omega$~Centauri (image credit: ESO, https://www.eso.org/public/images/eso0844a/). The WFM data in the middle panel consists of 10 pointings of $1\arcmin\times1\arcmin$ each that were repeatedly observed for at least~$>12$~epochs.   Two of the pointings were taken $\sim$5$'$ away from the cluster center to cover the half-light radius of the cluster. The right panel shows the NFM data which spans $\sim$20$''\times$15$''$ of which each pointing has a size of 7.5$''\times$7.5$''$.}
    \label{fig:pointings}
\end{figure*}

% \begin{figure*}[!t]
%     \centering
%     \includegraphics[width=1.1\linewidth]{figures/5139_voronoi.png}
%     \includegraphics[width=\linewidth]{figures/zoomed_5139.png}

%     \caption{Voronoi binning of the stars: Top: WFM and NFM data binned together with the bins consisting of at least 100 stars per bin. Bottom: Voronoi maps of NFM data, within 15$''$ of the cluster. Different centers from the literature are marked on the maps. Both the left panels are the velocities whereas right panels are the velocity dispersions. XXX These maps are still in progress 
%     }
%     \label{fig:voronoi}
% \end{figure*}

\section{Muse Data - Observation and Reduction}
\label{sec:data}
% \begin{table}
% 	\centering
% 	\caption{This is an example table. Captions appear above each table.
% 	Remember to define the quantities, symbols and units used.}
% 	\label{tab:example_table}
% 	\begin{tabular}{lccr} % four columns, alignment for each
% 		\hline
% 		A & B & C & D\\
% 		\hline
% 		1 & 2 & 3 & 4\\
% 		2 & 4 & 6 & 8\\
% 		3 & 5 & 7 & 9\\
% 		\hline
% 	\end{tabular}
% \end{table}
\begin{table*}
    \centering
    \caption{MUSE observations analysed in this work summarized per pointing.}
    \begin{tabular}{cccccccc}
    \hline
    Pointing no. & Mode & RA         & Dec       & $N_{\rm epochs}$ & $N_{\rm exposures}$ & Total exp. time & Median seeing \\
                 &      &            &           &                  & $[{\rm s}]$     & $[{\rm arcsec}]$ \\ \hline
    Pointing 01  & WFM  & 13:26:45.0 & -47:29:09 &    15 & 45 & 2025 &   0.62 \\
    Pointing 02  & WFM  & 13:26:45.0 & -47:28:24 &    14 & 42 & 1890 &   0.61 \\
    Pointing 03  & WFM  & 13:26:49.5 & -47:29:09 &    17 & 51 & 2253 &   0.60 \\
    Pointing 04  & WFM  & 13:26:49.5 & -47:28:24 &    17 & 51 & 2295 &   0.62 \\
    Pointing 05  & WFM  & 13:26:40.6 & -47:28:31 &    14 & 46 & 3646 &   0.60 \\
    Pointing 06  & WFM  & 13:26:53.9 & -47:29:01 &    17 & 51 & 4080 &   0.62 \\
    Pointing 07  & WFM  & 13:26:36.8 & -47:27:54 &    15 & 45 & 4500 &   0.60 \\
    Pointing 08  & WFM  & 13:26:31.0 & -47:29:55 &    16 & 49 & 7350 &   0.62 \\
    Pointing 11  & WFM  & 13:26:40.3 & -47:25:00 &    14 & 43 & 12900 &   0.92 \\
    Pointing 12  & WFM  & 13:26:41.0 & -47:24:03 &    12 & 36 &21600 &   0.82 \\
    Pointing 91  & NFM  & 13:26:47.2 & -47:28:50 &     2 & 8 & 4800 &   0.08 \\
    Pointing 92  & NFM  & 13:26:46.1 & -47:28:45 &     2 & 8 & 4800 &   0.08 \\
    Pointing 93  & NFM  & 13:26:46.8 & -47:28:46 &     3 & 11 & 6080 &   0.06 \\
    Pointing 94  & NFM  & 13:26:46.5 & -47:28:51 &     2 & 8 & 4800 &   0.07 \\
    Pointing 98  & NFM  & 13:26:46.6 & -47:28:49 &     1 & 4 & 2400 &   0.08 \\
    Pointing 99  & NFM  & 13:26:47.5 & -47:28:51 &     1 & 4 & 2400 &   0.07 \\ \hline
    \end{tabular}

    \label{tab:observations}
\end{table*}

The observations used in this paper were carried out as a part of MUSE guaranteed time observations (GTO) of 25 GCs (PI: S. Kamann, S. Dreizler). We selected the clusters to be within 15 kpc and have central velocity dispersions of~$>$5~km/s. They were observed during multiple epochs,  enabling binary stars to be detected. An overview of the survey is provided in \citet{kamann18}.  Another survey was also carried out using the MUSE-General Observer (GO) program, 105.20CG.001
(PI: N. Neumayer) and published in \citet{nitschai23}, which contains the catalog with a combination of the GO data and our GTO data. For the analysis of $\omega$ Centauri here, we only used a combination of wide field mode (WFM) and narrow field mode (NFM) GTO observations. 
%The pointings are shown in Figure~ \ref{fig:pointing}.

The WFM observations consist of the data presented in \citet{latour21} plus three additional epochs of observations taken during the nights 2021-05-09, 2021-08-05, and 2022-05-29 as part of program 105.20CR and 109.23DV. In total, 10 WFM pointings of $1\times1\,{\rm arcmin}$ each have been observed for 15 epochs on average.  To this dataset, we add NFM observations consisting of 6 pointings of size $7.5\times7.5\,{\rm arcsec}$, observed during 6 nights (2019-04-06, 2019-05-03, 2020-02-23, 2021-05-09, and 2022-05-29) as part of observing programs 0103.D-0204, 0104.D-0257, 105.20CR, and 109.23DV.  An overview of the different pointings used in this work, including their number of epochs, total exposure times, and median seeing values, is provided in Table~\ref{tab:observations}.  The location of NFM pointings 91 to 94 was chosen to cover the different centers of $\omega$~Centauri that have been proposed in the literature \citepalias{NGB08,N10,vdMA10}, while pointings 98 and 99 resulted from a misidentification of the requested tip-tilt star at the telescope.  VRI mosaics created from the reduced WFM and NFM data are shown in Figure~\ref{fig:pointings}.  The NFM data spans $\sim$20$''\times$15$''$, while the WFM data spans approximately 3$\times$5~arcmin other than the two pointings that are $\sim$5~arcmin away from the center. These pointings were observed to complete the radial coverage inside the half-light radius of the cluster.

The raw data were reduced using the standard MUSE pipeline \citep{weilbacher2020} in versions 1.2 and later.  All NFM data were reduced with pipeline version 2.8, which includes a strongly improved NFM flux calibration compared to the previous versions.  We used the default settings of the pipeline, with two exceptions. First, we did not perform a sky subtraction, as it would also remove stellar light given the crowding of the observed fields. Second, we did not perform a correction for telluric absorption, which instead was corrected during the analysis of the spectra (see below).  Data cubes were created for individual pointings and epochs, and they typically combined three (for WFM observations) or four (for NFM) exposures.  In between exposures, derotator offsets of 90~degrees and small spatial dithers were applied.

The reduced data cubes were processed using \textsc{PampelMuse} \citep{kamann13,pampelmuse}, which performs a deblending of the individual stellar spectra based on a wavelength-dependent model of the point spread function (PSF) that is recovered from the data and a wavelength-dependent coordinate transformation from a reference source catalogue to the MUSE data.  As reference catalogues for $\omega$~Centauri, two publicly available \textit{Hubble} space telescope (HST) data sets were used. The central WFM pointings (01-05, cf. Table~\ref{tab:observations}), as well as all NFM pointings used the catalogue created by \citet{anderson08} for the ACS survey of Galactic GCs \citep{sarajedini07}.  The outer WFM pointings (06-12), which are not or only partially covered by the ACS footprint, used the photometric catalogue generated by \citet{anderson10}.  As a PSF model, we used a Moffat profile for all WFM cubes and optimized the FWHM as well as the kurtosis \citep[parametrized by $\beta$, cf.][]{kamann13} as a function of wavelength.  In cases where visual inspection of the cubes suggested elongated stars, we included the ellipticity and the position angle (PA) of the semi-major axis of the Moffat in the set of free parameters.  While the Moffat profile has been shown to accurately describe the WFM-PSF \citep{fusco20}, it cannot describe the more complicated shape of the NFM-PSF.  Hence, the NFM data were instead processed using the \textsc{Maoppy} model developed by \citet{maoppy}, which has previously been successfully applied to NFM observations of the Galactic globular cluster M80 (NGC~6093) by \citet{goettgens21}.

The extracted spectra were processed in several analyses to measure the LOS velocities and determine the stellar parameters of the corresponding stars.  These analyses rely on useful initial guesses, which we obtained by comparing the aforementioned photometric catalogues to isochrones from the database of \citet{bressan12}, where we assumed an age of $13~{\rm Gyr}$, a metallicity of $[{\rm Fe/H}]=-1.33$, and an extinction of $A_V=0.37$.  Initial guesses for the surface temperature $\log g$ and the effective temperature $T_{\rm eff}$ of each star were obtained by finding the point that is closest to the isochrone in a color-magnitude diagram (CMD).  A $m_{\rm F606W}-m_{\rm F814W}$ vs $m_{\rm F606W}$ CMD was used for the \citet{anderson08} catalogue, while a $m_{\rm F435W}-m_{\rm F625W}$ vs $m_{\rm F625W}$ CMD was used for the \citet{anderson10} photometry. Note that despite the strong evidence for metallicity and/or age variations within $\omega$~Centauri, we only used a single isochrone when deriving initial parameter guesses.  This is because all parameters except for $\log g$ were later refined during the spectral analysis. $\log g$ was not refined because it impacts the shapes of the spectral lines rather than their strengths, and thus is difficult to measure at the low spectral resolution of MUSE. Moreover, $\log g$ is only weakly dependent on the age and metallicity of the old stars present in the cluster. 
% The reason why $\log g$ is not refined spectroscopically is that it impacts the shapes of the spectral lines rather than their strengths, and thus is difficult to measure at the low spectral resolution of MUSE.
\begin{figure*}
    \includegraphics[trim={0cm 0cm 0cm 0cm},clip,width=\linewidth]{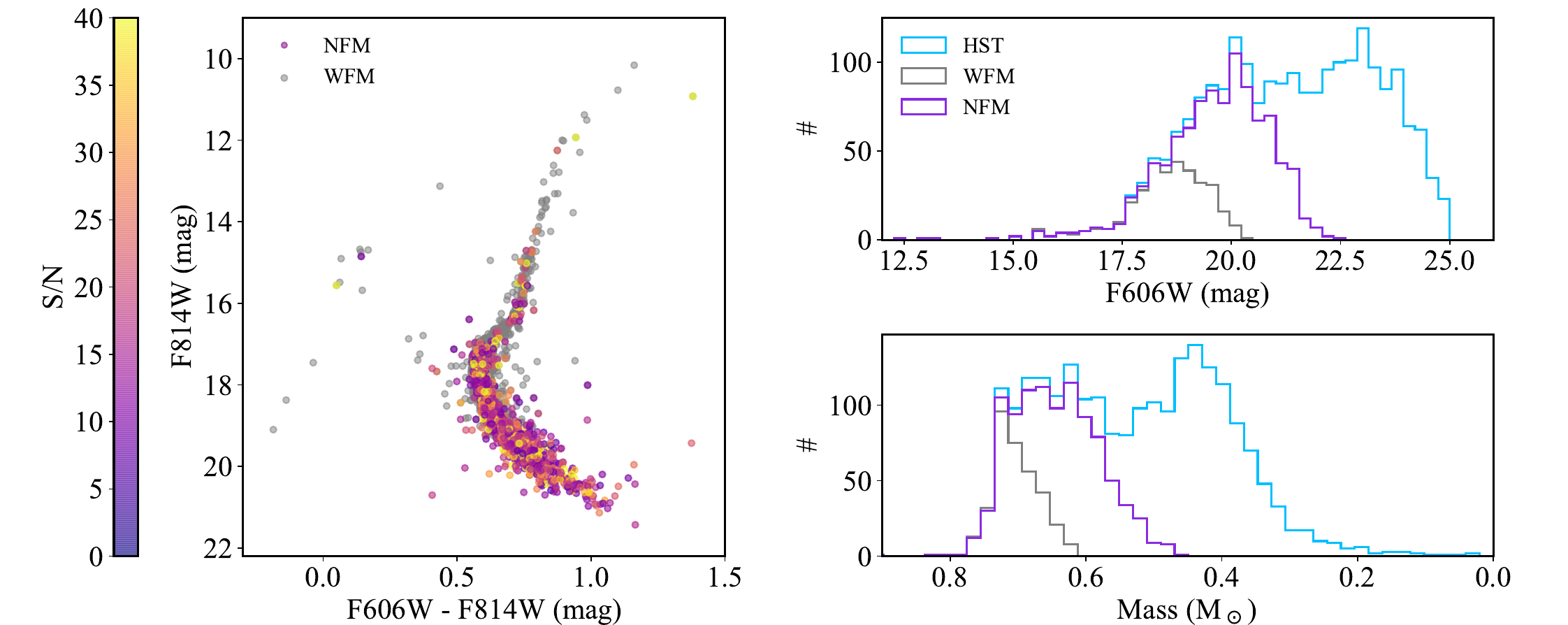}
     \caption{Left:  Color magnitude diagram of $\omega$~Centauri. The points are the 963 NFM (colored) and 1552 WFM (gray) stars within 20$''$ from the \citet{vdMA10} center. They are colored by the S/N of the spectrum as determined during the extraction of the MUSE spectra. Right:  Completeness histogram of MUSE NFM and WFM data as a function of magnitude and stellar mass on the NFM footprint. Gray lines correspond to the WFM data and the solid purple lines represent NFM data. The NFM data goes 2 magnitudes deeper  than the WFM data and extends beyond 20 mags in the F625W band within 20$''$ from the center. The mass range of the stars covered by the NFM is also broader with stellar masses starting from $\sim$0.5~M$_\odot$.
    }
    \label{fig:cmd}
\end{figure*}
To infer LOS velocities, we first cross--correlated each extracted spectrum against various templates drawn from the PHOENIX spectral library presented in \citet{husser13}.  The templates were chosen to represent the range in stellar parameters expected in $\omega$~Centauri ($T_{\rm eff}=\{3\,000~\text{K}, 6\,000~\text{K}, 9\,000~\text{K}\}$, $\log g=\{1, 3, 5\}$), but had solar metallicity.  We selected the LOS velocity provided by the template that gave the strongest cross--correlation signal and used it as an initial guess for the following full-spectrum fit with \textsc{Spexxy} \citep{husser16}. When reliable initial guesses could not be obtained from the cross--correlation, the spectra were then initialized at the systemic velocity of $\omega$~Centauri.  During the full spectrum fit, which was performed using templates from the PHOENIX library of \citet{husser13}, we determined $[{\rm Fe/H}]$ and $T_{\rm eff}$ alongside the LOS velocity.  As illustrated in \citet{husser16}, \textsc{Spexxy} enables the user to fit the telluric absorption simultaneously with the stellar spectrum, which was done while analysing the spectra.  To verify the wavelength accuracy of every MUSE cube, we determined the average velocity of the telluric absorption fits of all spectra extracted with an ${\rm S/N}>30$.  This mean telluric velocity, which usually varied between $-1$ and $1\,{\rm km/s}$, was subtracted from all velocity measurements obtained for the cube. 

The derived stellar velocities were assessed based on a reliability flag following the method described in \S~3.2 of \citet{giesers19}. This includes considering the signal-to-noise ratios of the extracted spectra or the agreement between the velocities derived from cross--correlation and spectral fit. Only velocities with a reliability grading above 80\% were kept. Further, we considered the agreement between the magnitudes recovered from the spectra and those provided in the HST catalogs, as large discrepancies could indicate contamination in the spectra by nearby stars. \textsc{PampelMuse} expresses the photometric agreement as a \textit{MagAccuracy} parameter, ranging from 0.0 (large discrepancy) to 1.0 (perfect agreement). We imposed a cut value of 0.6.
% This is based on several factors such as checking for reasonable uncertainties, S/N of the spectrum, results from \textsc{Spexxy} that are plausible, etc. All of these criteria are then combined to a reliability parameter and only the velocities that have a reliability above 90\% are used. The velocities per star were averaged after the filtering, using the inverse-variance weighting.  To filter the velocities we used those stars that were not contaminated by other stars, within the field of view, and a magnitude accuracy of at least 0.6 mags. 

The final step in the data analysis was a calibration of the uncertainties of the LOS velocity measurements.  To do so, we followed the approach outlined in \citet{kamann18}, which makes use of the different valid velocity measurements available for each star and is based on the expectation that in the absence of intrinsic variability, the normalized scatter of these measurements should follow a Gaussian distribution with a standard deviation of unity. Following this calibration, we averaged the filtered velocities per star, using inverse-variance weighting.

 After extracting the reliable velocities, 38,602 stars were left in our sample, including the NFM and WFM data. To compare the depth of the NFM data, we plot stars within 20$''$, which include 963 NFM stars and 1552 WFM stars, on the color-magnitude diagram (CMD) of Figure~\ref{fig:cmd}.  We used the ACS photometry for the CMD, which has photometry in the F606W and the F814W filters. The completeness histograms show the comparison of number counts in the WFM and NFM samples relative to the HST catalogue within the NFM region. We found that the NFM data is complete for stars brighter than 19.5 mags, which excludes $\sim$10 stars that are on the edges of the pointings in these comparisons. For the WFM data, we are complete below 18 mags. 50\% of the stars are below 18.6 mags in WFM mode whereas, in the NFM data, 50\% of the stars are below 19.8 mags, which is close to the completeness limits. The mass range covered by the NFM data is also broader since more stars with lower masses (down to 0.5~M$_\odot$) are included. The masses of the stars were obtained based on the isochrone comparisons as mentioned above. There are no observations for the WFM data below 20 mags, whereas the NFM data extends down to the 22nd magnitude. But, further away from the center, the WFM data reach a similar depth due to longer exposure times and lesser crowding. The dataset covers everything from the tip of the RGB to 4 mags below the main-sequence turnoff. The median S/N is $\sim$12 for the NFM data and the median velocity uncertainties are $\sim$2.3~km/s. For stars around 16 mags, the typical uncertainties in velocities are $\sim$1~km/s, whereas, for a star of magnitude $\sim$21 mags, the uncertainties increase to $\sim$5~km/s because of the low S/N. The 50\% completeness limit in the F606W magnitude and the masses are 18.6 mags and 0.71~M$_\odot$ respectively for the WFM data. For NFM data, the 50\% completeness limit is 19.8~mags and 0.65~M$_\odot$.

% For the line of sight towards each cluster, we generated a realization of the contaminant population using the Besanc¸on model of the Milky
% Way (Robin et al. 2003). The available photometry was used to
% constrain the list of simulated stars to the apparent magnitude range
% of our data. The simulation provided us with an expected distribution of the non-member stars in the radial velocity–metallicity
% plane towards each cluster. The expectation–maximization method
% was then used to compare the measurements of [M/H] and vr of every star to the probability densities expected for the cluster and the
% non-member stars. Each star was assigned a probability of cluster
% membership such that the overall likelihood was maximized under
% the boundary condition that the fraction of cluster to non-member
% stars decreases monotonically with increasing distance to the cluster centre. For more details on the method, including the formulae
% that have been used, we refer to Walker et al. (2009) and Paper II.

The membership probability of the stars being a part of the cluster was estimated using the procedure described in \citet{kamann18}. 
 Briefly, we assume that the observed stars are composed of cluster members and a field population. For the former, we assume that its velocity and metallicity distributions can be approximated as Gaussians, whereas for the latter, we use the Milky Way model by \citet{robin03} to predict velocities and metallicities. We then iteratively determine the likelihood of each star belonging to the cluster or field population, with the additional constraint that the fraction of cluster stars decreases monotonically with distance to the cluster center.
% {\bf It is based on a model developed by \citet{robin03} for the Milky Way, wherein, first a contaminant population is simulated. Then, an expected distribution of non-member stars with radial velocities and metallicities are compared to the measurements obtained using the expectation-maximization method. The likelihood of a {\bf star being cluster member} or a foreground star is maximized using the condition that the fraction of cluster to non-cluster stars decreases as we go radially outward from the cluster center.  
The membership probability is then assigned to each star and those with a probability lower than 0.6 are excluded from our analyses. Although $\omega$~Centauri has a  metallicity spread of more than a factor of 10 \citep[e.g.,][]{johnson10}, it has a radial velocity that offsets the member stars from those of the MW stars. For a more detailed description of the method, see \citet{walker09} and \citet{kamann16}. We also excluded the stars with temporal variations in their LOS velocities that can be potential binary stars. We use the probabilities that were derived by using the method described in \citet{giesers19}. This work will be presented in Wragg et al. \textit{(in prep.)}.

The final sample of stars that were used in the analysis of the central kinematics excludes non-member stars with a cluster membership probability cut of~$<$0.6, a binary variability probability~$>$0.7, and an S/N cut of~$<$8 for the mean of the extracted spectra, which leaves us with 28,108 stars of which 936 are NFM stars.

% To exclude non-member stars and those with low S/N, we employ a S/N cut of $>$8 and a cluster membership probability cut of $>$0.6, which leaves us with 27,134 stars of which 739 are NFM stars. 

% \section{Analysis of kinematic data}
% We create radial profiles of the kinematic data by binning according to the distance to the cluster center. Each radial bin consisted of at least 100 stars. Then we used Voronoi binning code \citep{cappellari03} to create 2D kinematics around the cluster centre. The Voronoi bins consist of at least XX stars per bin.

\section{Kinematic Data}
We analyse the LOS velocities in this section within the half-light radius of $\omega$~Centauri and create velocity and velocity dispersion radial profiles. This will help further study of a possible IMBH in $\omega$~Centauri. The center of $\omega$~Centauri is debated in the literature and several centers have been proposed based on photometry and kinematics. 

\subsection{Existing Analyses and Proposed Centers}

A putative BH in $\omega$~Centauri was reported in \citetalias{NGB08}, and the center was estimated using the surface density counts of stars in the central 40$''$ of HST/ACS data. The kinematic data was obtained from GMOS integral-field spectroscopy with a spectral resolution of R~=~5560~${\text \AA}$ in the Ca-Triplet region (7900 - 9300~${\text \AA}$), and the LOS velocities were derived using these spectra. Their dataset was seeing-limited, whereas  the NFM data is at a resolution of 25~mas. The field-of-view (FOV) is also bigger covering the half-light radius of the cluster. Based on the LOSVD measurements from GMOS IFU and a clear cusp in the surface brightness profile, an IMBH of M$_\mathrm{BH}$~=~4.0$^{+0.75}_{-1.0}\times$10$^4$~M$_\odot$, and a LOS central velocity dispersion, $\sigma_\mathrm{LOS}$ = 23.0$\pm$2.0 km/s were reported for this cluster.  The measured cusp in the surface brightness profile was in agreement with the theoretical predictions from $N$--body simulations in \citet{baumgardt05} for clusters harboring IMBHs, where the central BH tends to prevent the core collapse.

\citetalias{vdMA10} presented another analysis for the central few arcmins, where they estimated the projected number density distribution of $\sim$10$^6$ stars from HST photometry. The LOS velocities were obtained using the ground-based data from \citet{suntzeff96,mayor97,reijns06}. Proper motions were also estimated using ground-based data from \citet{vanleeuwen00} and also using 10$^5$ stars from HST data \citep{anderson10}. The center of the cluster was determined using three independent ways and found to be offset from the one derived in \citetalias{NGB08} by 12$''$. Their analysis did not confirm the density cusp that was observed in \citetalias{NGB08} due to the offset center. Anisotropic models were fit to the data wherein if a flat core was assumed, a no-BH model provided a good fit, whereas cuspy models required either an IMBH of M$_\mathrm{BH}$~=~8.7$\pm2.9\times$10$^3$~M$_\odot$ or a dark cluster of size $\lesssim$~0.16~pc. The final result was an upper limit on a possible IMBH with M$_\mathrm{BH} \lesssim$~1.2$\times$10$^4$~M$_\odot$ at 1$\sigma$ confidence level, in strong tension with the IMBH mass suggested by \citetalias{NGB08}.

A third search for a central IMBH was performed in \citetalias{N10} where they obtained additional integrated field spectroscopy for the central region of the cluster using VLT-FLAMES/ARGUS. These observations have a spectral resolving power of R$\sim$10000 and cover the 820 -- 940 nm region using the GIRAFFE spectrograph with a FOV of 11.5$''\times$7.3$''$. They combined the new LOS velocities derived from the Ca-Triplet region with the existing measurements from \citetalias{NGB08} and estimated a different center from the additional kinematics. A peak in the velocity dispersion map was found by running a 5$''$ kernel across the 2-D map of velocity dispersion, which was their estimated center for the cluster along with a measured central velocity dispersion of 22.8$\pm$1.2~km/s. Using isotropic dynamical models and their kinematic center, they estimated a central BH of M$_\mathrm{BH}$~=~4.7$\pm$1.0$\times$10$^4$~M$_\odot$. 

\begin{figure*}
    \centering
    \includegraphics[width=0.95\linewidth]{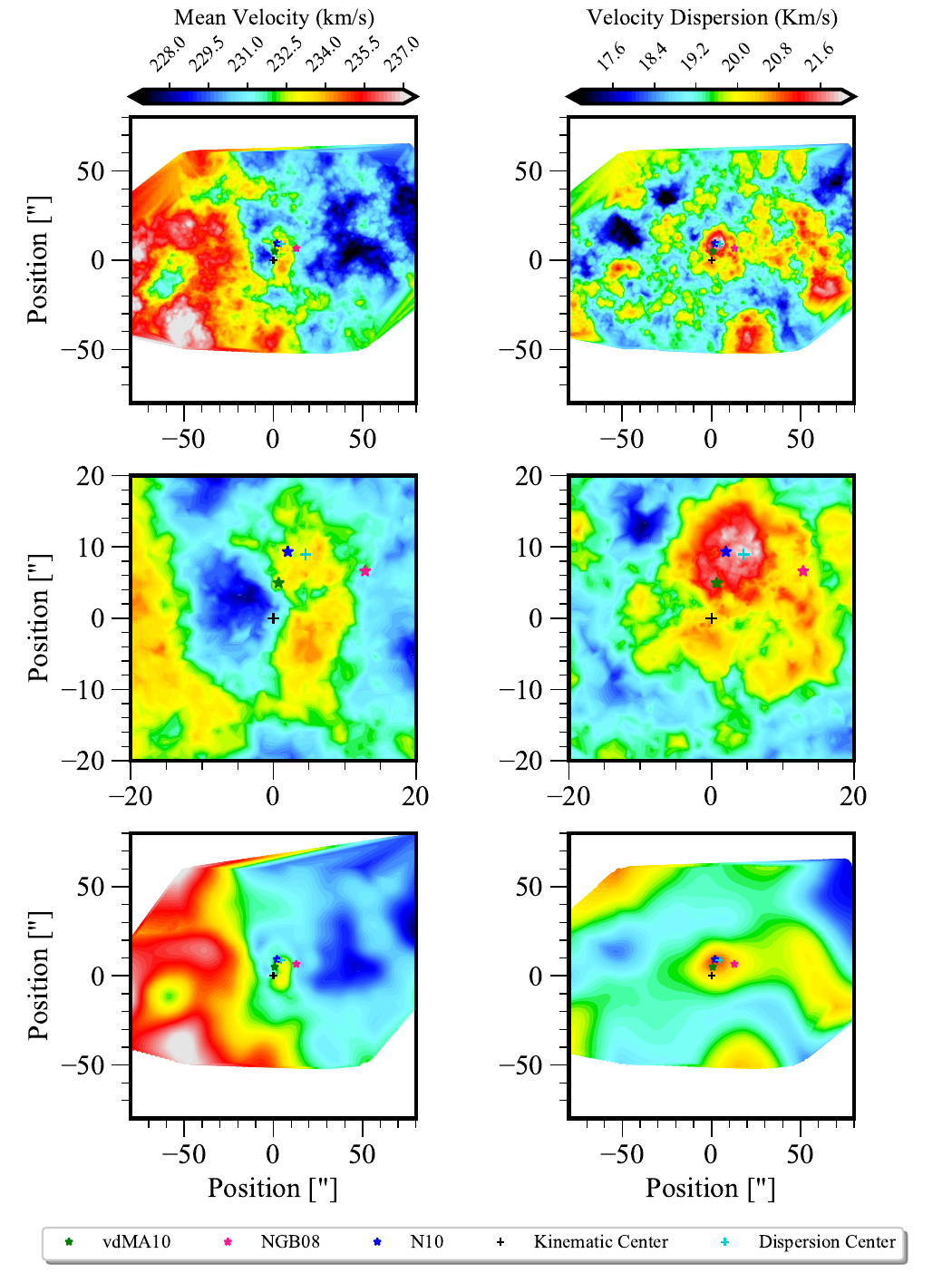}

    \caption{Rotation and velocity dispersion maps: \textit{Top \& Middle Panel}: Rotation velocity and velocity dispersion maps of the central 80$''$  and a zoomed version of the central 20$''$ of the MUSE WFM \& NFM kinematics grouped using KNN with a k-value of 300.
    \textit{Bottom panel}: LOESS smoothed 2-D maps of the rotation and dispersion. Different centers are marked in different colors as indicated in the legend.
 }
    \label{fig:clustercore}
\end{figure*}

The above-mentioned measurements using various centers are listed in Table~\ref{table:threecenters}. The \citetalias{NGB08} center is offset by 12$''$ from the \citetalias{vdMA10} center. The \citetalias{N10} and \citetalias{vdMA10} seem to lie approximately on the rotation axis of the cluster but are $\sim$3$''$ apart (see Figure~\ref{fig:clustercore}). Further analyses using these centers are presented in the subsequent sections. 

\begin{table}
    \centering

\caption{Existing and current measurements of $\omega$~Centauri}
\label{table:threecenters}
\def\arraystretch{1.2}
\setlength{\tabcolsep}{2pt}
% \begin{threeparttable}
\begin{tabular}{ccccc}
\hline\hline
  &Center RA &  Center Dec & M$_{BH}$ (M$_\odot$) & $\sigma_{LOS}$ (km/s)	\\
\hline	
\citetalias{NGB08} & 13:26:46.04 & -47:28:44.8 &  4.0$^{+0.75}_{-1.0}\times$10$^4$ & 23.0$\pm$2.0 \\ 
\citetalias{vdMA10} & 13:26:47.24 & -47:28:46.5 & $\lesssim$1.2$\times$10$^4$ & -- \\
\citetalias{N10} & 13:26:47.11 &  -47:28:42.1 & 4.7$\pm$1.0$\times$10$^4$ & 22.8$\pm$1.2 \\
Kinematic Center & 13:26:47.31 & -47:28:51.4 & -- & 19.3$\pm$1.4 \\
(From Rotation) & && & \\
Dispersion Center &13:26:46.86 & -47:28:42.5& -- &22.6$\pm$1.5 \\
(From Dispersion) &&&& \\
% Center RA &13:26:46.043    & 13:26:47.24  & 13:26:47.1125 \\
% Center Dec        & -47:28:44.8  & -47:28:46.45   & -47:28:42.074 \\
% M$_{BH}$ (M$_\odot$)   &  4.0$^{+0.75}_{-1.0}\times$10$^4$  & $\lesssim$1.2$\times$10$^4$ & 4.7$\pm$1.0$\times$10$^4$ \\ 

% $\sigma_{LOS}$ (km/s)  & 23.0$\pm$2.0    & --  &22.8$\pm$1.2  \\

\hline

\end{tabular}

% \begin{tablenotes}
$Note$: The first three rows list the existing measurements of $\omega$~Centauri made in \citetalias{NGB08} \citep{NGB08}, \citetalias{vdMA10} \citep{vdMA10}, and \citetalias{N10} \citep{N10}. The bottom rows present measurements derived in this work, which are described in \S~\ref{sec:kinematic_center}. 
% \end{tablenotes}
% \end{threeparttable}
\end{table} 

\subsection{Analysing the MUSE Kinematics}
\label{sec:methods}
To study the dynamics of $\omega$~Centauri, we created radial profiles of the velocity dispersion and the rotation velocity using two methods. In the first method, we radially binned the stars, where each bin consisted of a minimum of 100 stars, and covered a radial range of $\log (r/1'')>$~0.15. In the second method, we created analytical profiles for the cluster to estimate the rotation and line-of-sight velocity dispersion. 
For both methods, a maximum likelihood approach \citep{pryor93} was used in combination with the Markov chain Monte Carlo (MCMC) algorithm \textsc{emcee} \citep{foreman13} to estimate the quantities. A detailed description of this analysis is given in \citet{kamann18}. For the maximum likelihood analysis, we assumed the probability of finding a star with a velocity $(v_i\pm\epsilon_i)$ at a projected distance $r_i$ from the cluster center to be: 
\begin{equation}
\label{eq:probability}
    p(v_i,\epsilon_i,r_i) = \frac{1}{2\pi\sqrt{\sigma_r^2+\epsilon_i^2}} \exp{\frac{(v_i - v_0)^2}{-2(\sigma_r^2+\epsilon_i^2)}}
\end{equation}
 Here, $\sigma\mathrm{_r}=\sigma\mathrm{_r}(r_i)$ and $v_0=v_0(r_i)$ are the dispersion and heliocentric radial velocity of the cluster, respectively, at the position of star $i$. We then found the values of $\sigma_{\rm r}$ and $v_{\rm 0}$ that minimized the negative log-likelihood of the model given the kinematic data. A limitation of this approach is that it applies to the stellar systems whose LOSVD is Gaussian ignoring the higher moments of the LOSVD. 
 For determining the rotation velocity of the cluster, we assume that the cluster is a rotating disc and add an angular dependence ($\theta$) to the mean velocity in Equation~\ref{eq:probability}. 
\begin{equation}
    \label{eq:vel0}
    v_0 \rightarrow \overline{v}(r_i,\theta_i) = v_0 + v\mathrm{_{rot}}(r_i)\sin(\theta_i-\theta_0(r_i))
\end{equation}
where $v_{\rm rot}$ and $\theta_0$ are the projected rotation velocity and axis angle.
This allows us to estimate the LOS velocity in each bin radially from the cluster center. We then implemented the MCMC analysis to minimize the negative likelihood in every radial bin and ran a total of 100 chains with 100 steps for each bin, which is sufficient for determining the parameters. We constrained three parameters, $\sigma_r$, $v\mathrm{_{rot}}$, and $\theta_0$ with uniform, uninformative priors.

% Due to the binning of a small number of stars, it is hard to see the rotation that is observed close to the center. It is better visible in the radial profiles of the cluster. 

 Since the effect of binning is removed while using the analytical forms, we create the analytical profiles
% and dispersion profiles,
using the following functional form for the rotational velocity, v$\mathrm{_{rot}}$:
 \begin{equation}
 \label{eq:vrot}
  v\mathrm{_{rot}}(r) =\frac{2v\mathrm{_{max}}r}{r\mathrm{_{max}}} \bigg/ [1+(r/r\mathrm{_{max}})^2]
  \end{equation}
  % \begin{equation}
     % \sigma\mathrm{_r}(r) = \sigma\mathrm{_{max}}\big/[1+(r/a_0)^2]^{0.25} 
  % \end{equation}
The v$_\mathrm{rot}$ profile is based on the prediction for a system undergoing violent relaxation \citep{lynden-bell67} and has been previously used to model the rotation profiles of star clusters in, e.g., \citet{bianchini18}. 

% and $\sigma_{r}$ is characterized by a Plummer profile \citep{plummer1911}.
 Here three parameters were constrained, the peak amplitude of the rotational velocity- v$_\mathrm{max}$, the radius at which v$_\mathrm{max}$ occurs- r$_\mathrm{max}$,
% ,the scaling factor for Plummer profile- $\sigma\mathrm{_{max}}$, the Plummer radius- a$_0$, 
and  $\theta_0$. Using the MCMC analysis as mentioned above, we fitted those parameters for a total of 100 chains and 500 steps and created radial profiles based on the median values and the 16th and 84th percentile of the distributions. The results of these methods applied to the data are discussed and presented in  \S~\ref{sec:comparison} and Figure~\ref{fig:profile}.

% In the following Section we analyse this counter-rotation of the central 20$''$ of the cluster. 

% \begin{figure*}
%     \centering
%     \includegraphics[trim={2.5cm 2cm 6.5cm 3cm},clip,width=\linewidth]{figures/disp_loess.pdf}

%     \caption{Velocity dispersion maps of $\omega$~Centauri. The maps describe the velocity dispersion determined using the K-nearest neighbors method. From left to right the k--value increases from 40 to 300 in the KNN analysis. A central peak is visible for the k--value of 300. This peak coincides with the center from \citetalias{N10}, since their center was derived based on the peak of the velocity dispersion. 
%  }
%     \label{fig:disp}
% \end{figure*}

% \begin{figure*}
%     \centering
%     \includegraphics[width=0.6\linewidth]{figures/delta_center.pdf}
%     \caption{Center estimate from kinemetry, when the center was allowed to vary for the ellipses. This was performed on the LOESS map from Figure~\ref{fig:clustercore}.
%     }
%     \label{fig:center_estimate}
% \end{figure*}
\section{A counter-rotating core in $\omega$~Centauri}
In this section, we analyse the central 100$''$ in detail with a focus on the central 20$''$ as this area is covered by the NFM data. 

\begin{figure*}
    \centering

    \includegraphics[trim={0cm 0cm 0cm 0cm},clip,width=\linewidth]{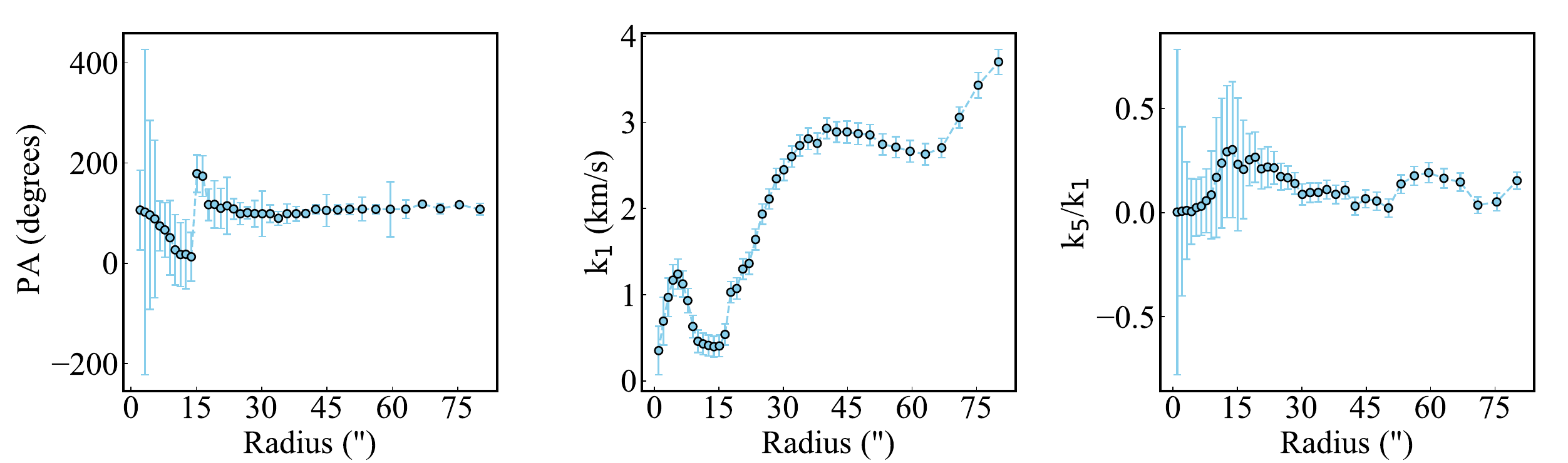}
    \caption{Kinemetry results for the central 80$''$ of the cluster using the LOESS maps. Left: The PA of the cluster,  measuring the change in the velocity structure on the map. Middle and Right: The coefficients k$_1$ and k$_5$/k$_1$ from the Fourier expansion of the velocity field for the best-fitting ellipse. k$_1$ describes the bulk motion of the cluster with a peak at $\sim$5$''$ and decreases after that. Then it increases with the radius indicating the global rotation of the cluster. k$_5$/k$_1$ is peaked at $\sim$15-20$''$, which indicates the rotation components within the map.}
    
    \label{fig:rotation_amp}
\end{figure*}

\subsection{Creating Kinematic Maps}
\label{sec:maps}
 To study the mean LOS velocity as a function of position, we used a supervised neighbors-based learning method to create a smoothed velocity map. We used a K-nearest neighbors (KNN) analysis, where the basic principle is to group the stars based on a metric \citep{pedregosa11}. Here, the grouping classifies the nearby neighbors based on their distances and we estimated those using  \texttt{biweight\_location} \citep{beers90} of \textsc{astropy}, which performs a robust detection of the mean of the distribution, in this case, the LOS velocities of the stars.

In this method, the number of samples ($k$) is defined as the number of neighbors that would be chosen to represent a group. A smaller $k$--value selects fewer neighbors and results in a coarser grouping whereas a higher $k$--value provides smoother grouping. At first, standard Euclidean distances are calculated between the neighbors based on the $k$--value, and then a maximum radius is given within which the grouping needs to be performed. We visually inspected maps of the KNN-based mean velocities with $k$--values ranging from 40--300.  We found that smaller values of $k$~($<$200) lead to patchiness of the velocity maps due to the lesser number of stars in each group (for example, see Figure~\ref{fig:disp}). Therefore, we used a $k$--value of 300 to derive the LOS velocity map, where the large-scale rotation of $\omega$~Centauri is visible (see Figure~\ref{fig:clustercore}, left column). A zoomed-in version shows the central 20$''$ of the core (Figure~n\ref{fig:clustercore},left middle panel), mainly covered by the NFM data, which is counter-rotating with respect to the global rotation of the cluster. The central 100$''$ of $\omega$~Centauri are rotating with maximum projected velocities of $\sim$4~km/s and the inner core within 20$''$ shows maximum rotational velocities of $\sim$3--5~km/s. The different centers from the literature are marked here, with each center being a few arcsec apart. Other estimates of the centers based on our velocity and dispersion maps are also marked, which we discuss in the next subsection. 
% Our estimate of the center is closest to the \citetalias{vdMA10} center with a distance of 5$''$ between them.

 For a better visualization of the velocity maps with less noise, we employed another method to derive the velocity map of the cluster; the locally weighted regression (LOESS) technique, which was introduced in \citet{atlas3d20} based on an algorithm developed by \citet{cleveland79} for the 1-D case and further improved for the 2-D case in \citet{cleveland88}. This is a regression method that uses a multivariate smoothing analysis on a surface by the local fitting of the function in a moving way similar to the moving average. This method robustly determines the mean values of the underlying parameters in case of noisy data. In our case, the parameters are the LOS velocities and the noise is the velocity dispersion of the stars. We used the code \texttt{loess\_2d} from \citet{atlas3d20}, which requires input velocities, and a regularization factor (f) that describes the fraction of points that are considered for the approximation and controls the amount of smoothing of the map. We used a value of f~=~0.1 and assumed a linear approximation to determine the LOESS smoothed maps. The bottom panel of Figure~\ref{fig:clustercore} shows the LOESS map of the LOS velocities within 80$''$ around the cluster center. The result is consistent with the maps from the KNN analysis. On a larger scale, the global rotation is dominant and as we go towards the center, the counter-rotation is visible. The different centers are marked similarly as in the KNN maps, and we observe that except for the \citetalias{NGB08} center, the rest of the centers are aligned close to the rotation axis of the counter-rotating core. It is worth noting that the \citetalias{NGB08} center is also close to the zero velocity curve (green contour) of the counter-rotation, but on the other side, which is a local minimum.
% Further analysis of the centers based on this LOESS map is discussed in the following sections.

In both maps, there is clear evidence that the rotation in the central region of the cluster is misaligned with the rotation at large radii. This counter-rotation dominates the kinematics within the central 10--20$''$ and appears to be centered south of the \citetalias{vdMA10} and \citetalias{N10} centers.
% at approximately RA= 13h 26m 47.31s and Decl = -47$^\circ$28$'$51.39$''$ .
To check for the effect of contamination from any bright stars, we removed stars brighter than 14th magnitude in the F625W filter. We found that there are two bright stars at the centers of the counter-rotating velocity components both moving in the same direction with a velocity of $\sim$20~km/s but the counter-rotation signal is still significant even after removing the stars. 

Apart from the LOS velocities, we also analysed the velocity dispersion of $\omega$~Centauri using KNN analysis. Here we grouped the stars and used the statistic \texttt{biweight\_scale} from \textsc{astropy}, which determines a robust standard deviation of the distribution of stars, i.e. a velocity dispersion in this case. We also used different $k$--values ranging from 40 -- 300. A lot of peaks or patchiness was detected across the entire velocity dispersion map for smaller $k$--values (see Figure~\ref{fig:disp}). But, as we grouped more stars, for a $k$--value of 300, we found a prominent central peak. This is shown in the right column of Figure~\ref{fig:clustercore}.  This peak is close to the center from \citetalias{N10} as expected since their center was derived based on the peak in the velocity dispersion. The remaining centers are also marked similarly to the ones in the velocity maps. The bottom-most panel shows the LOESS smoothed map that was derived based on the KNN velocity dispersion map, where the central peak is visible. When we compare the mean velocity and the velocity dispersion maps in the middle panel, the peak in the central velocity dispersion appears to be offset from the rotation axis of the counter-rotating core. We quantify this rotation and dispersion in the subsequent sections and constrain different centers based on them. 

% This is also visible in our velocity dispersion profiles that are described later in \S~\ref{sec:comparison}. 

% To derive a center based on the peak in the velocity dispersion is ideally not reliable since the kinematic center need not coincide with the photometric center of the cluster. The dispersion profile will trace the peak in the map, thus rising towards the center and can support the detection of a possible black hole. In $\omega$~Centauri, the center in \citetalias{vdMA10} was based on the number density counts of the stars, which is offset from the velocity dispersion peak and the \citetalias{N10} center by 3$''$. We used the counter-rotation to constrain a center for this cluster. 
% The centers are marked similarly as in the KNN maps,  

% (XXX SKA: This signal should also be discussed with regards to Fig. 3, i.e., is the is the rotation field consistent with what is observed in Fig. 3?)

\subsection{Quantifying the counter-rotation with Kinemetry}
\label{sec:kinemetry}
To quantify the velocities and rotation, we performed \textit{kinemetry} on the velocity and dispersion maps of the cluster. We used LOESS maps for this analysis only as an alternative to the KNN maps to test if the counter-rotating core is real and quantifiable using kinemetry. \textit{Kinemetry} is a method from \citet{krajnovic06} that performs a harmonic expansion of 2-D kinematic moments, in our case, mean local velocities that are the first kinematic moments, along a set of best-fitting ellipses on the map.

When analysing velocity maps, kinemetry assumes that there are ellipses along which the velocities can be described with a simple cosine law, 
\begin{equation}
\label{eq:vel}
    V(R, \psi) = V_0 + V\mathrm{_c}(R)\sin(i) \cos(\psi),
\end{equation}
where V$_0$ and V$_c$ are the systemic and circular velocities, projected on the sky at an inclination $i$, and traced along the ellipse via the azimuthal angle $\psi$, measured from the projected major axis. Eq.~\ref{eq:vel} is strictly correct for disks in which stars move on circular orbits, but \citet{krajnovic08,krajnovic11} showed that it applies to a large fraction of early-type galaxies, with deviations of less than 5\%. The kinemetric analysis of the velocities along an ellipse is performed by evaluating the harmonic terms: 
\begin{equation}
    K(a,\psi) = A_0(a) + \sum_{n=1}^{N} A_n(a) \sin(n\psi) + B_n(a) \cos(n
    \psi)
\end{equation}
where a and $\psi$ are the semi-major axis and the azimuthal angle, and in the case of the velocity, n is an odd number. The best-fit ellipse is determined by minimizing the A1, A3, and B3 terms, as they define the shape and the orientation of the ellipse. 
Kinemetry results are often presented in a compact form: 
\begin{equation}
    K(a,\psi) = A_0(a) + \sum_{n=1}^{N} k_n(a) \cos([n(\psi - \phi_n(a))]
\end{equation}
where $k_n = \sqrt{A_n^2 + B_n^2}$ and $\phi_n =$ arctan $(A_n/B_n)$.
In this form the term k$_1$ describes the amplitude of the rotation, while k$_5$ is the first non-minimized higher order coefficient which defines the deviations from a simple rotation as assumed in Eq.~\ref{eq:vel}. For further details regarding \textit{kinemetry}, we refer to \citet{krajnovic06}.

For the kinemetric modeling, for the initial models, we fixed the center of the cluster to an initial guess from the previous section. Since the photometric axis ratio of $\omega$~Centauri is above 0.8 at all radii \citep{geyer83}, we limited the shape (axial ratio) of the ellipses for the kinemetry to be between 0.7 and 1.0. To increase the range of the fit, we used a cover value of 0.6, which means that 60\% of the points on the ellipse should be present for the ellipse to be included in the fit. We also fixed the range for the PA from 0$^\circ$ to 180$^\circ$. The results from these fits are shown in Figure~\ref{fig:rotation_amp}. 

The left panel of Figure~\ref{fig:rotation_amp} shows the PA that is varying with respect to the radius and twists at $\sim$15--20$''$ by 180$^\circ$ approximately indicating the observed counter-rotation. The errors on the PA of the data points below 5$''$ are large and can be ignored because the counter-rotation is on a scale of 15$''$ and the ellipses fitted on a scale smaller than 5$''$ are unable to trace the rotation. The k$_1$ coefficient tracing the rotation shows a small peak at 5$''$ delineating the extent of the counter-rotation followed by the global rotation of the cluster. The large error bars on k$_5$, due partially to the very low-level rotation of the cluster, make it almost consistent with 0 and prohibit further analysis of the deviations. The peak in k$_5$ at $\sim$20$''$ is correlated with the drop in k$_1$, as is expected at the edge of a kinematic component \citep{krajnovic06}. Together with the radial variation of the PA of ellipses (Figure~\ref{fig:rotation_amp}, left), which stabilizes at the end of the counter-rotating core, k$_1$$\sim$0, and where the error in k$_5$ significantly drops, kinemetry analysis quantifies the extent and the shape of the central counter-rotation in $\omega$~Centauri. To constrain the center, we explore the central 20$''$ with different techniques in the next section.

% The center is found to be the closest to that in \citetalias{vdMA10}.

\begin{figure*}
    \centering
         \includegraphics[trim={0cm 5 0 0},clip,width=0.75\linewidth]{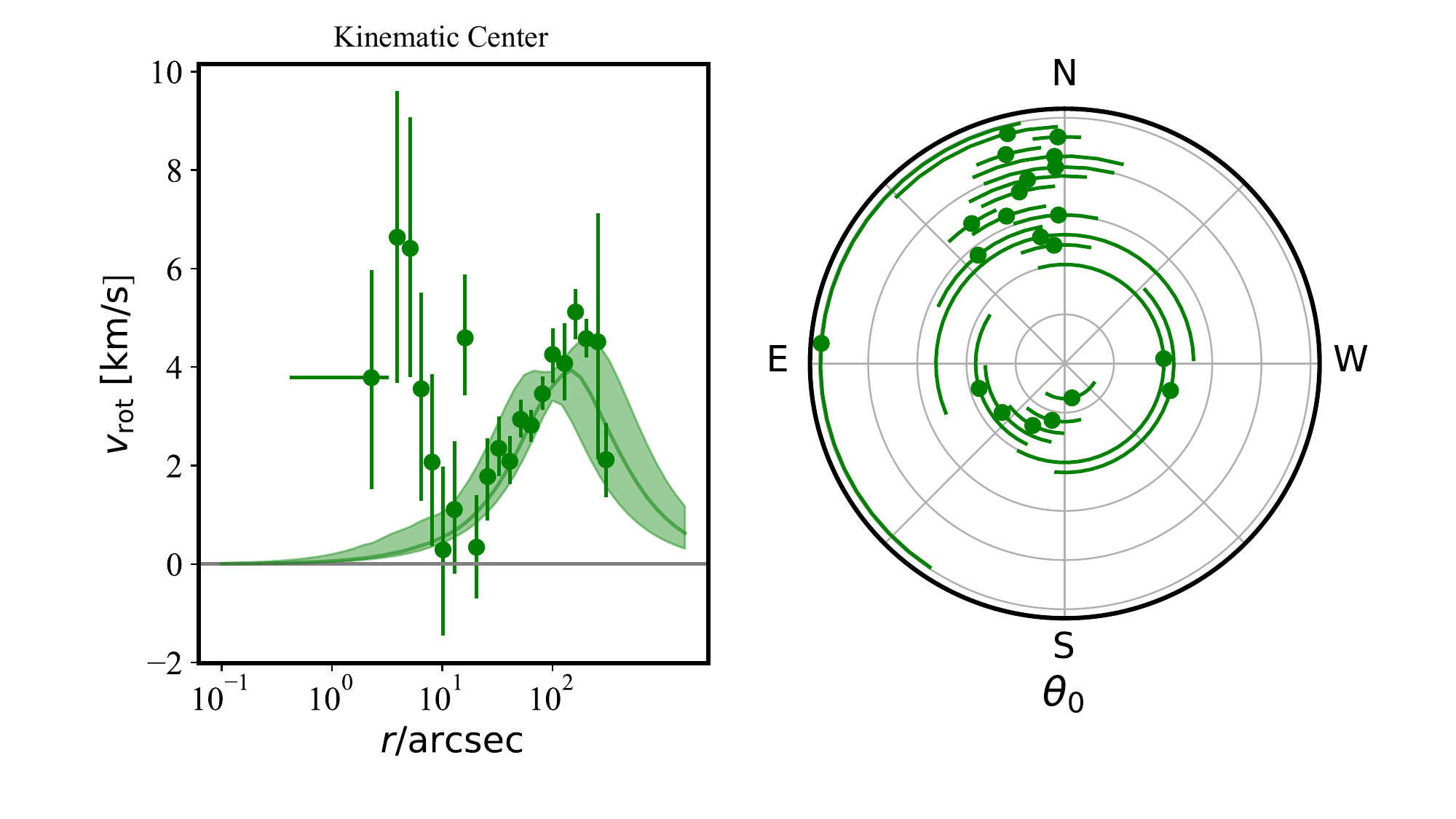}
    \caption{Left: Rotation profile of the cluster using the kinematic center. Solid points are the result of discrete binning of the radial velocities (Eq.~\ref{eq:vrot}) whereas the shaded curve is the analytical rotation profile Right: PA of the rotation of the cluster. Note the counter-rotation observed towards the center.}
    \label{fig:profile}
\end{figure*} 

\begin{figure*}
    \centering
    \includegraphics[width=\linewidth]{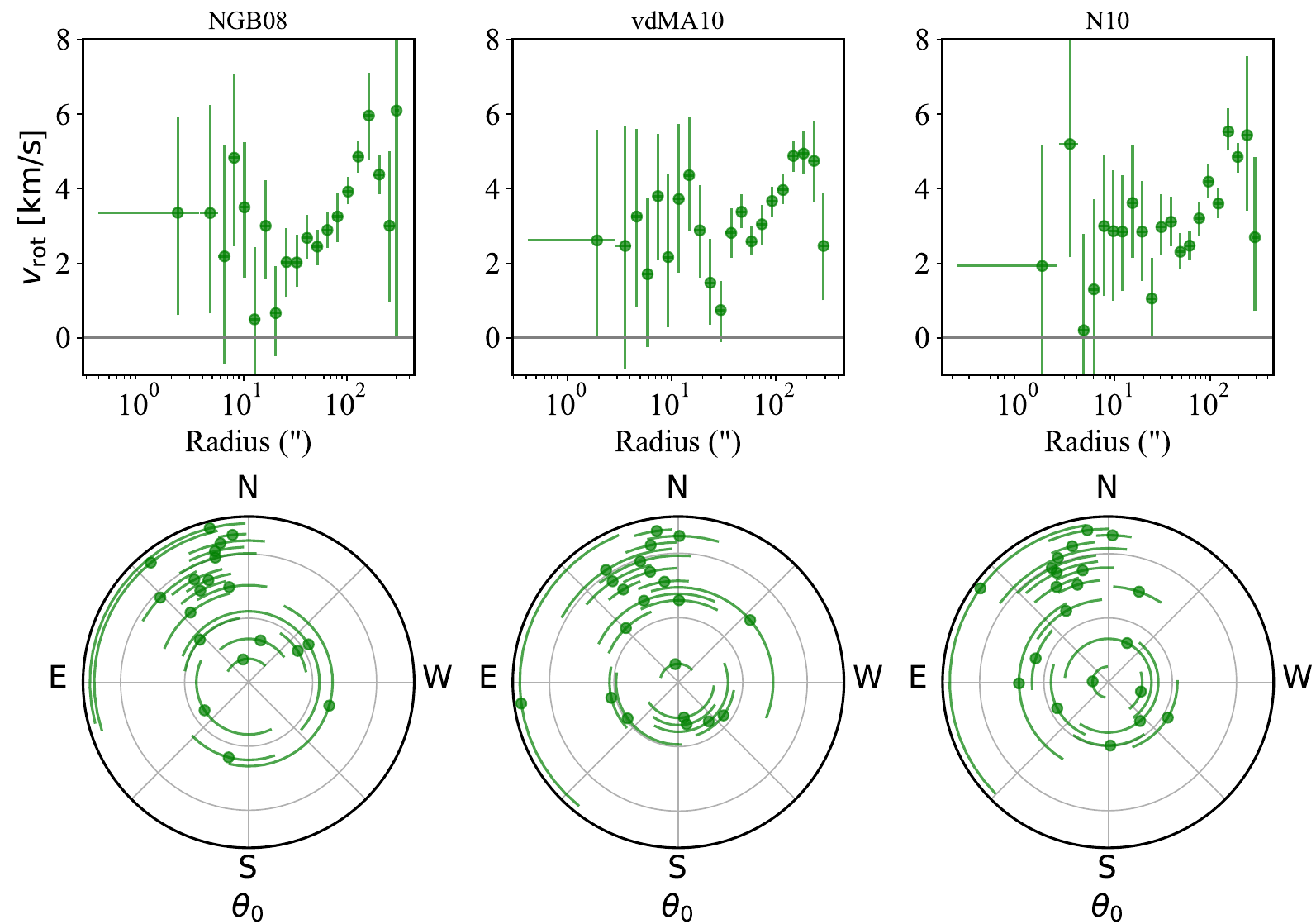}
    \caption{ Rotation profile and PA of the rotation of $\omega$-Cen. The top panels show the rotation (amplitude) profile of the stars that are binned radially using the method in \S~\ref{sec:methods}. The bottom panels show the orientation of the rotation axis within each radial bin. From left to right: profiles using \citetalias{N10}, \citetalias{vdMA10}, and \citetalias{NGB08} centers. Note that the counter-rotation is only observed in \citetalias{vdMA10}.
    }
    \label{fig:radial_profiles}
\end{figure*}

 \section{Kinematic Center and comparisons with the previous studies}
In this section, we describe various methods that were used to determine the center for this cluster and then use it to compare it with the previously published centers.
\subsection{Finding the Kinematic Center}
\label{sec:kinematic_center}

A well-defined center is crucial in determining the velocity dispersion and rotation profiles, which are required for the detection of any kind of dark mass in the cluster center. Below, we show that using several methods, we are unable to numerically constrain the center to better than $\sim$5$''$, within the range of previous estimates. The kinematics near the center are so complex that the result also depends on the method we use. However, assuming that the counter-rotation represents the center, we constrain the center to be at RA=~13h~26m~47.31s and Decl~=~-~47$^\circ$28$'$51.39$''$ with an uncertainty of $\sim$5" along the declination.  This position is closest to \citetalias{vdMA10}, while the centers \citetalias{N10} and \citetalias{NGB08} centers are not consistent with being at the center of the counter-rotating core.  We describe each of the methods attempted to better constrain the center below; all of these are consistent with our center quoted above, with uncertainties ranging from 5$''$--20$''$.

% We use the MCMC simulations on the kinemetric analysis to determine the center. We created LOESS map and then created a grid of centers. then ran mcmc and got the center with 

The first method we implemented was a slight modification to the MCMC analytical profile fits described in \S~\ref{sec:methods}. Here, we added two additional parameters to the MCMC analyses, dx and dy as offsets to the center and fit them in the iterations. We provided an initial guess for the center by eye based on the counter-rotation from the velocity map of the KNN analysis and allowed it to vary as offsets in dx and dy.
We fit the center for every radial bin and based on several runs, we found that the best-fit median of the center was not very well constrained. Although dx and dy are scattered around zero, the average errors ranged up to 20$''$, which is our entire region of counter-rotation. To test the accuracy of this method, we used a sample test case of M80 using the data published by \citet{goettgens21}, where the center was  constrained using a similar MCMC analysis of their Jeans' models. The dx and dy offsets that we obtained were similar to those found in \citet{goettgens21}, with median errors up to 2$''$. We conclude that this method is most likely not suitable for $\omega$~Centauri, due to the flatness of the core and the counter-rotation present close to the center.

We tried an approach using the pie-slice method, a similar approach that was used in \citet{anderson10}, where they used the number counts of the stars to determine the center whereas we use the LOS velocities instead of the photometry. This method determines the point around which the stellar kinematics are most symmetrically distributed. First, we provide the same initial guess that was used in the previous method as a trial center. Based on this center, a 5$''$ grid of trial centers is created around it. Around each grid point, we constructed 16 azimuthal wedges, extending to distances of 20$''$, and each wedge was further divided into 7 radial bins. We minimized the differences between the mean velocities in opposite pairs of bins along with the uncertainties using the \texttt{biweight\_location} robust statistic (described in \S~\ref{sec:maps}). Finally, we summed up these differences for all the bins to determine a $\chi^2$ measure. This method yielded a $\chi^2$ minima along the rotation axis of the counter-rotation of the cluster and hence appears not suited for locating the center of $\omega$~Centauri. 

% We used an approach based on kinemetry, where we kept the center as a variable parameter. We provided an initial guess for the center and allowed the center to vary in our fits along with the best-fitting ellipse. We found that the estimate for the center scatters around the initial guess but the uncertainties on the center estimate vary with the radius of the cluster. The central position has a standard deviation of 5$''$. The uncertainties increase strongly ($\sim$40$''$) as we go close to the radius of counter-rotation. We ignore the center estimates from these ellipses as these are at the edge of the counter-rotation. After a radius of $\sim$20$''$, the errors are in the range of $\sim$10$''$, but the center estimate starts to deviate here and for the outermost bins, the center is well constrained. This is mostly because the outer ellipses are constraining the center for the global rotation instead of the counter-rotation at the center. To mitigate this issue, we focus only on the central 30$''$.
% e ignore the center estimate from the outer ellipses as they trace the global rotation. Within the inner 20$''$, the estimate for the center varies around the initial guess, but with uncertainties ranging from 5$''$--10$''$. 

Finally, we used an approach based on kinemetry, where we performed a grid search using the center as two variables. We divided the central 20$''$ into a grid of 30 centers and performed kinemetry using  each center. Each center was fixed in the code along with the PA and the flattening (q) of the ellipses. To obtain the initial values for PA and q, we first ran the kinemetry code with our initial guess for the center and averaged the PA and q of the ellipse at $\sim$3$''$ and $\sim$5$''$. We used this value PA = 80$^\circ$ and q = 0.84 for fixing the PA and q of the ellipses. To avoid the effects of the global rotation, we limited the number of ellipses to 9. and then performed kinemetry on the grid of centers. We estimated the $\chi^2$ for the last three ellipses by minimizing the sum of the squares of their coefficients A1, A2, B2, A3, and B3 \citep{Jedrzejewski87,krajnovic06}. We found that two local minima existed for the $\chi^2$ along the counter-rotating axis that lies $\sim$1.5$''$ and $\sim$7$''$ from our initial guess. We used a mean of both the centers to get an estimate for the final center with a final uncertainty estimate of $\sim$5$''$ in the declination and $\sim$1.5$''$ in the right ascension. The uncertainty was estimated using the 1$\sigma$ map of the $\chi^2$ for a $\Delta \chi^2$ = 2.3. Hereafter, we refer to this center as the kinematic center.

Based on this result and our earlier approaches to determining the center, we found that the location of the center is always degenerate along the rotation axis. The \citetalias{vdMA10} center that lies along this rotation axis  has the potential to be one of the possible centers as it is the closest to our derived center and is within the median uncertainties of 5$''$. 

We also estimated a center based on the LOESS dispersion maps from Figure~\ref{fig:clustercore}. For using kinemetry on the dispersion maps, the profile is assumed to be constant and the corresponding coefficients are minimized. In this case, the dispersion maps are assumed to be point symmetric (refer \citet{krajnovic06}), and since we are not analysing the rotation, we used circles instead of ellipses with a PA of 0$^\circ$. We then minimized the sum of the square of the coefficients A1, A2, B1, B2, A3 and B3 to obtain the $\chi^2$. Here, A$_0$ provides an estimate of the velocity dispersion, which is 22.6$\pm$1.5~km/s. We similarly performed kinemetry as previously mentioned and used the $\chi^2$ estimate to determine the center. We estimated the center to be RA~=~13h~26m~46.86s and Decl~=~-47$^\circ$~28$'$~42.46s with an uncertainty of 1.5$''$. Hereafter, we refer to this center as the dispersion center. The uncertainty is estimated similarly as determined previously using the $\chi^2$ map. This dispersion map based center is closest to the \citetalias{N10} center and is  $\sim$2.5$''$ away from it. Although, there is a significant offset ($\sim$10$''$) between the counter-rotation center and the dispersion center, both the centers are closest to the \citetalias{vdMA10} and \citetalias{N10} centers respectively. On the contrary, the offset in the \citetalias{NGB08} is the largest. All the centers and corresponding central dispersion values are listed in Table~\ref{table:threecenters}.

\subsection{Kinematic Profiles for the Various Centers}
\label{sec:comparison}
% The center that we determine using the velocity maps is closest to the \citetalias{vdMA10} center, which is $\sim$5$''$ away. The \citetalias{vdMA10} center, \citetalias{N10}, and the kinematic center lie close to each other (within 8$''$) and along the axis of the observed counter-rotation. The \citetalias{NGB08} center is the furthest and $\sim$13$''$ away from the kinematic center.
\begin{figure*}
    \centering
    \includegraphics[width=0.9 \linewidth]{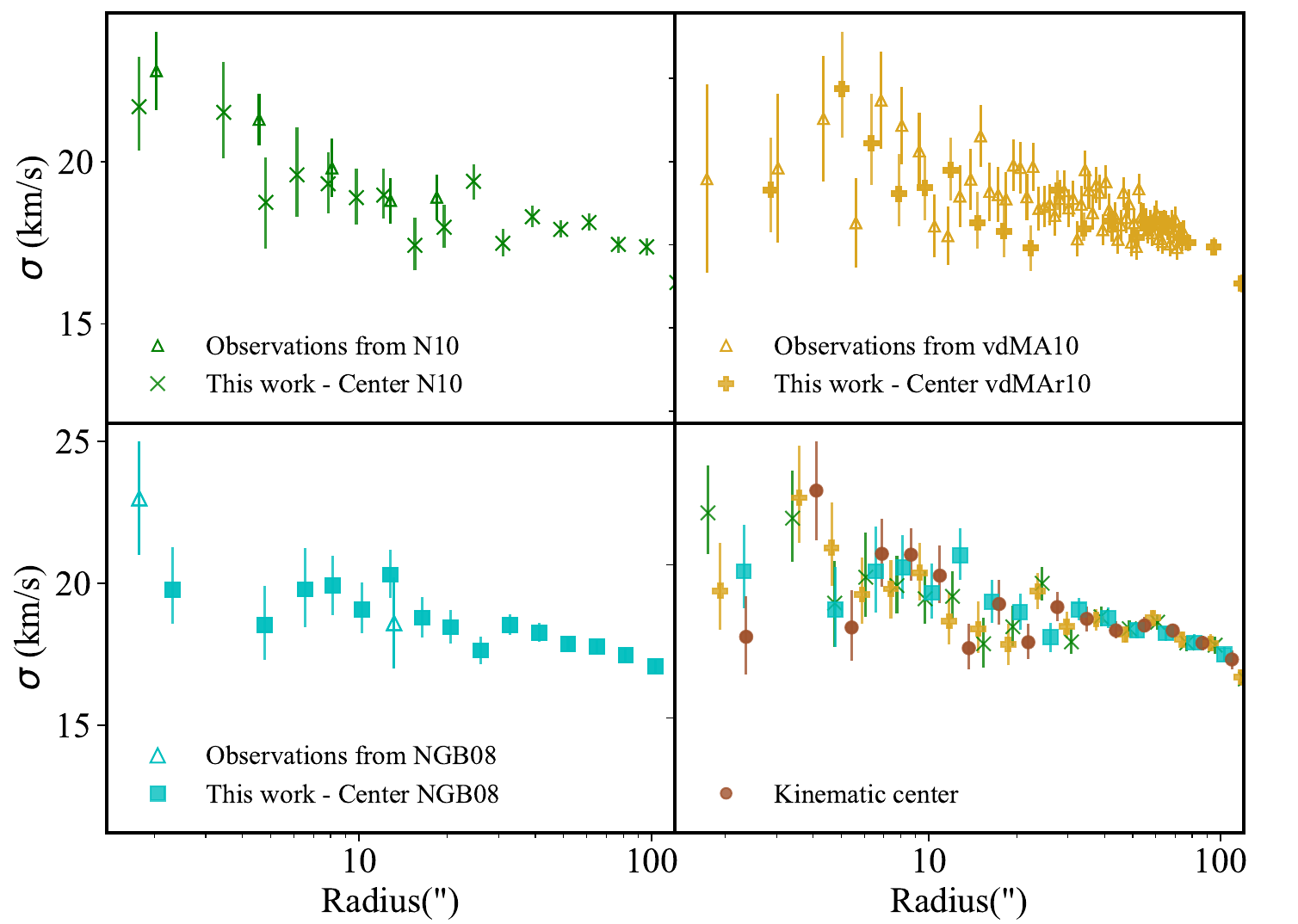}
    \caption{Comparison of the velocity dispersion profiles. Open triangles in green, orange, and blue are observations from the \citetalias{NGB08}, \citetalias{vdMA10}, and \citetalias{N10}. Other symbols in green, orange, and blue are the radially--binned velocities from this work for different centers. Note that our measurements match the overall dispersion profiles well from each study. The bottom right panel shows the velocity dispersion profile (brown solid circles) derived using the kinematic center (center from the counter-rotation) in this work. The other three profiles are the same as shown in the rest of the panels and are plotted for comparison with the kinematic center's profile.}
    
    \label{fig:comp_disp}
\end{figure*}
We choose our derived center from the counter-rotation i.e. kinematic center for the rest of the analysis. We do not use the center from dispersion in our analyses as it is close to the \citetalias{N10} center and results in similar profiles.
 We used the kinematic center to derive the rotational velocity profile, dispersion profile, and the PA of the rotation as shown in Figure~\ref{fig:profile} and Figure~\ref{fig:comp_disp}. These profiles were derived using the method described in \S~\ref{sec:methods}. The solid points in Figure~\ref{fig:profile} are a result of the discrete binning of the radial velocities whereas the curve is the analytical profile of the rotation (from Eq.~\ref{eq:vrot}). In the rotation plot of the cluster, we see the signal for a counter-rotating core, where the PA of the velocities close to the center (within 20$''$) is in the opposite direction to the global rotation of the cluster.  The PA for the inner 15$''$ has a mean of 136$^\circ$ E of N with a scatter of 20$^\circ$ whereas the PA~$>$~20$''$ is $\sim$10$^\circ$. The maximum rotational velocity of the inner 15$''$ is 3-5~km/s whereas the outer rotational velocity peaks at $\sim$4.5 km/s.

 We also created radial profiles of the rotational velocity and velocity dispersions for the existing centers proposed in \citetalias{NGB08}, \citetalias{vdMA10}, and \citetalias{N10} using the same method. They are shown in Figure~\ref{fig:radial_profiles} and Figure~\ref{fig:comp_disp}. We observe that there is no significant counter-rotation around the \citetalias{NGB08} center. This is because the center is $\sim$12$''$ away from the other centers and further away from the counter-rotating core (see Figure~\ref{fig:clustercore}). This is also visible in the PA plot where the innermost data points (within 10$''$) are scattered around in all directions, whereas the rest show a global rotation for the cluster $\sim$19$^\circ$ E of N. The rotation curve around the \citetalias{vdMA10} center again shows a scatter close to the center without any distinctive peak but the PA plot shows a clear counter-rotating signal, which is similar to what we find for the kinematic center. The velocities within 10$''$ have a mean PA of 208$^\circ\pm$ 21$^\circ$ E of N except for the central data point and the rest of the velocities have a PA similar to the one from the previous center with a mean of 10$^\circ\pm$15$^\circ$. The velocities within 10$''$ from the \citetalias{N10} center are similar to those from the \citetalias{vdMA10} center but the PAs are scattered around in all directions, with some of the data points in the opposite direction compared to the global rotation. They have a mean of $\sim$200$^\circ$ but have a standard deviation of $\sim$60$^\circ$ which is much larger compared to the \citetalias{vdMA10} center.
The rotation profiles  at larger radii~($>$20$''$) for all the centers behave similarly as expected in GCs \citep[e.g.,][]{fiestas06}, including a detailed study of $\omega$~Centauri in \citet{vandeven06}. According to Eq.~\ref{eq:vrot}, we expect a maximum at a few half-light radii and then a steady decrease for larger radii. This trend was observed for 25  GCs in \citet{kamann18},  where detailed rotation and dispersion profiles were derived. We observe a similar trend, but in addition, there's also an increase in the rotational velocity close to the center in our estimates of $\omega$~Centauri due to the presence of counter-rotation in Figure~\ref{fig:profile}.

% The counter-rotation is close to the existing centers and could be an effect of an IMBH at the center. The sphere of influence for a BH mass estimate from \citetalias{N10} is $\sim$15$''$, which follows the rotation that we observe at the center. 

% The dispersion profiles also vary according to the center, as \citetalias{NGB08} shows almost a flat dispersion towards the center. The \citetalias{N10} dispersion profile clearly shows a rise towards the center, which was also observed in the original study, whereas the central data point drops down in the \citetalias{vdMA10}. The PAs are aligned to the axis of the rotation, which are the main indicators of rotation in this cluster. Within the central 20$''$, where we observe a strong counter-rotation in two of the three centers indicates that an accurate center determination is necessary for the dynamical analysis of this cluster. The analytical profiles, as expected are similar for all three cases indicating a flat dispersion profile towards the center, which is typical of a Plummer profile, with a dispersion value of $\sim$18.5~km/s.

Figure~\ref{fig:comp_disp} shows the comparison between the observed dispersion profiles derived for different centers.  First, we compared the dispersion profiles from the previously published data in \citetalias{NGB08}, \citetalias{vdMA10}, and \citetalias{N10} with the dispersion profiles derived using the MUSE data  for the same centers as used in those studies. The data from \citetalias{NGB08} consisted of only two data points from their observations using the GMOS-IFU. The measurements in \citetalias{N10} were done using the ARGUS IFU with FLAMES on VLT whereas the measurements for \citetalias{vdMA10} were taken from the HST proper motions that were estimated by \citet{anderson10}. We used a distance of 5.2~kpc \citep{harris96} for $\omega$~Centauri to scale the proper motions to the LOS velocity dispersions. 
We found surprisingly similar trends with respect to each study within the error bars. The \citetalias{N10} result has a rise in the central velocity dispersion that we observe from our data too; this favors the presence of a $\sim10^4$~M$_\odot$ BH from the \citetalias{N10} analysis.  The dispersion profiles from \citetalias{NGB08} and \citetalias{vdMA10} show a $\sim$11\% and $\sim$22\% drop for the center-most data point, respectively, compared to the \citetalias{N10} observations, but follow a similar trend in both profiles. Other than the \citetalias{N10} profile, all profiles seem consistent with a constant dispersion of $\sim$20~km/s in the central 10$''$, since from the velocity dispersion map, the centers are offset from the peak and lie close to the counter-rotation axis. 
The rightmost bottom panel of Figure~\ref{fig:comp_disp} includes the dispersion profile using our kinematic center along with the other dispersion profiles based on the MUSE data only. Including the uncertainties in the other profiles, the velocity dispersion rises smoothly towards the center up to a central velocity dispersion of $\sim$20~km/s, which is similar to other studies. There is no specific rise as observed in \citetalias{N10} though, which is mostly dependent on the assumed center.

 To quantify this rise in dispersion, we did a linear regression on the data within 15$''$ for the dispersion values using the code from \citet{kelly07}, which takes into account the measurement errors in the y-direction and estimates the scatter in the regression. We found variable slopes for each center with the highest slope of -3.9$\pm$1.8~km/s.arcsec for the \citetalias{N10} center. The slope for the \citetalias{vdMA10} center was found to be -3.2$\pm$1.9~km/s.arcsec. This is almost similar to the \citetalias{N10} center because the majority of the points rise but the central data point drops to 18~km/s. On the contrary, we found a zero to positive slope for the \citetalias{NGB08} center with a value of -0.3$\pm$3.1~km/s.arcsec. The slope for our kinematic center was -0.3$\pm$4.2~km/s.arcsec. From our regression analysis, we have a significant scatter for the slopes for each center especially for the \citetalias{NGB08} center and the kinematic center. Although they all were found to lie within 1$\sigma$ significance of each other, there are differences in the velocity dispersions within the central 5$''$ from each center. For the \citetalias{N10} center and the \citetalias{vdMA10} center, we estimated the integrated central velocity dispersions~($<$5$''$), which were 20.6$\pm$0.9~km/s and 20.9$\pm$0.7~km/s respectively. For the \citetalias{NGB08} center, it was 18.9$\pm$1.0~km/s whereas for the kinematic center, it was 19.6$\pm$0.9~km/s. The highest difference was found to be between the \citetalias{NGB08} center and the \citetalias{vdMA10} and \citetalias{N10} centers. The differences in these central velocity dispersions can indeed make a difference in the detection of an IMBH from the previous studies. A detailed dynamical modeling using the different centers and the dispersion profiles is needed to quantify the presence/absence of an IMBH.

 Another aspect that can be observed from the comparison of the dispersion profiles is that LOS velocity measurements from different types of observations such as integrated-light measurements or resolved stellar velocities result in similar velocity dispersion measurements. In the literature, such as \citet{luetzgendorf15}, significant changes between velocity dispersion measurements were found based on the type of observations. They compared the results from their previous work \citep{lutzgendorf11}, which was based on seeing-limited integrated light kinematics, and the results from \citet{lanzoni13} based on adaptive optics assisted resolved stellar kinematics in NGC~6388. They found that the integrated-light measurements can be biased towards higher velocities due to the contamination from bright stars close to the center. On the other hand, resolved velocities can also be biased because the contamination from neighboring stars can drive the individual star velocities to the mean velocity. We note that the previous observations from \citetalias{vdMA10} were using the HST proper motion data whereas the studies from \citetalias{NGB08} and \citetalias{N10} were using the integrated-light data from GMOS-IFU and VLT-FLAMES respectively. From our analyses, the velocity dispersions derived from discrete velocities are quite similar to the integrated-light measurements, and we do not find any notable difference in the measurements from resolved kinematics vs. the integrated-light measurements (done by other work). This implies that the contamination effects are small while extracting our kinematics. 
 Although the dispersion measurements are similar using different techniques, we note that the core density of  $\omega$~Centauri is log$\rho_c$~=~3.15~M$_\odot$/pc$^3$ whereas NGC~6388 has a log$\rho_c$~=~5.37 M$_\odot$/pc$^3$ \citep{harris96} which is significantly denser. Due to this, although the contamination of light from neighboring stars is higher for the integrated light measurements, the shot noise is much lower for NGC~6388 and is not affected by that uncertainty. For the MUSE data, we cannot make a straightforward comparison between the two clusters because of their different densities and different techniques involved in the extraction. Only a detailed analysis focusing on the differences between  extraction techniques can help identify if there is a significant difference in the measurements obtained for $\omega$~Centauri and NGC~6388.

% Although, note that there is no counter-rotation observed in the center from \citetalias{NGB08} (from Figure~\ref{fig:radial_profiles} in the \ref{app}).
% We plot them in a 1-D profile with respect to the radius in Figure~\ref{fig:rotation_amp}. We fix the rotation axis of the cluster, which is close to the major axis of the cluster and find the velocities and dispersion profile of the cluster using the same MCMC analysis as described in \S~\ref{sec:methods}. The peak rotation amplitude in the x-direction is then measured and is plotted in Figure~\ref{fig:rotation_amp}. Note that within the central 10$''$ it reverses the sign with a value of $\sim$5~km/s, and is a significant change. We also plot the position angle of the rotation, where it is allowed to vary freely. As expected, the PA clearly twists by 180$^\circ$ at a radius of $\sim$15$''$ indicating the presence of counter-rotation.
\section{high-velocity stars in $\omega$~Centauri}
A smoking gun for the presence of an IMBH would be the detection of high-velocity stars close to the center of $\omega$~Centauri using the NFM data. 
A discussion from \citet{baumgardt19} shows that for $\omega$~Centauri to host a 4.75$\times$10$^4$ M$_\odot$ BH, a tail of at least 20 high--velocity stars~$>$62~km/s  should exist within 20$''$ of the center (see Figure~2 from \citet{baumgardt19}). They ran a large grid of N-Body simulations that had several parameters such as the initial density profile, half-mass radius, the initial mass function, the cluster metallicity, and the mass
fraction of an IMBH in the clusters. In addition, they varied the retention fraction of the stellar-mass BHs. The best-fit model to the observed data had a 75\% retention fraction of stellar-mass BHs at the center and no IMBH. 

\begin{figure}
    \centering
    \includegraphics[trim={0.1cm 0.3cm 0.5cm 0.5cm},clip, width=\linewidth]{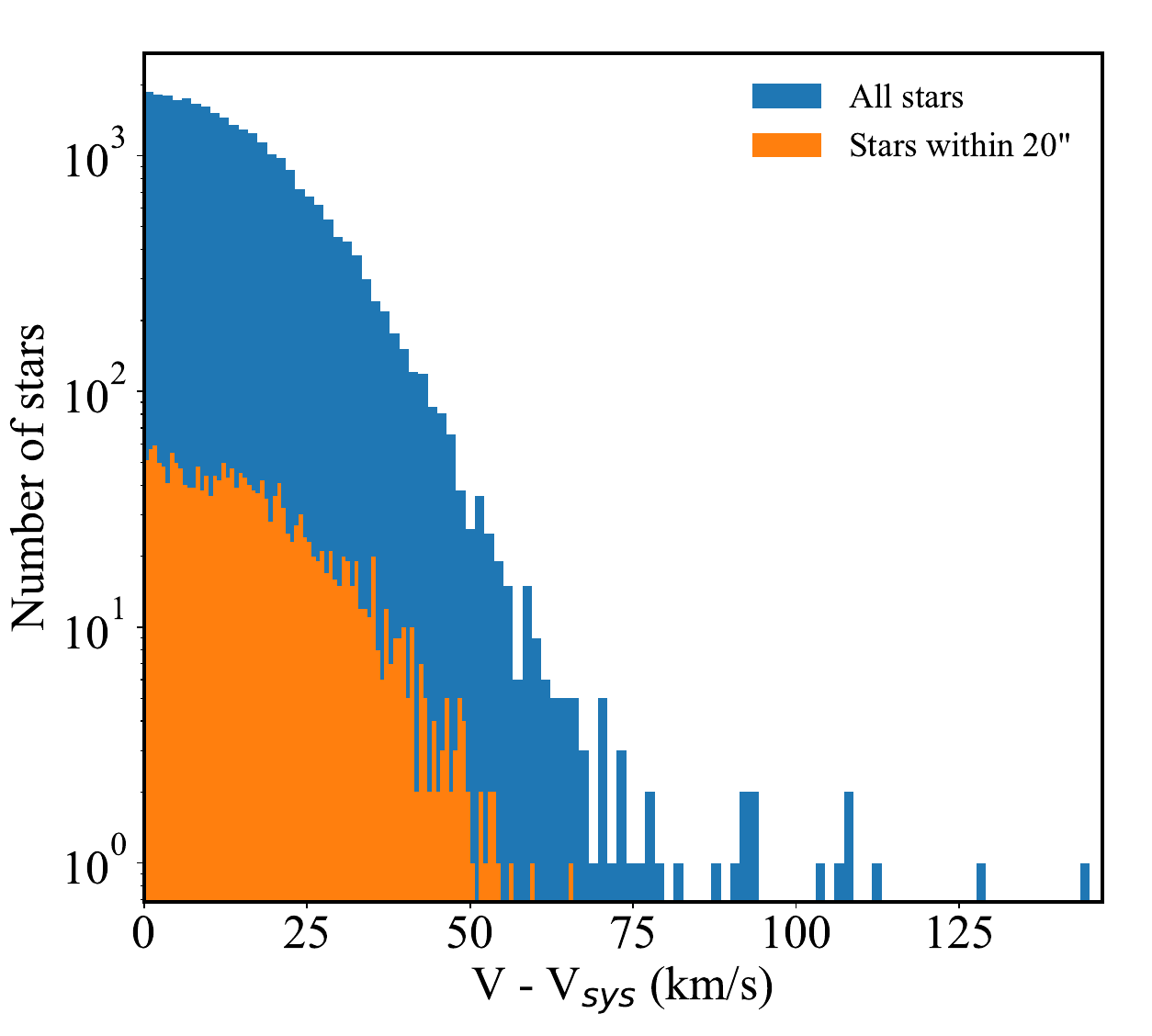}
    \caption{Histogram of the LOS velocities of all the stars with an S/N~$>$8 excluding the foreground stars. Note that we have a few high--velocity stars but they are $\sim$1 arcmin away from the center and none of them are close to the center (within 20$''$).
    }
    \label{fig:highvel_hist}
\end{figure}

To test this theory, we looked for evidence of any high-velocity stars from our entire sample of stellar LOS velocities. We use a S/N cut of 8 but we do not use a cluster membership cut, since the membership estimation is also dependent on the LOS velocity of the stars. After the cut, we had 28,782 stars, out of which we excluded stars that were obvious non-members and foreground stars of MW with a mean velocity close to zero. For this, we removed stars with absolute velocities~$<$75 km/s, which left us with a sample of 28,719 stars including members and non-members. We used our kinematic center to estimate the distances for this analysis. We plot all these stars in Figure~\ref{fig:highvel_hist}. This contains 50 stars that have a velocity~$>$62 km/s, but none are significant outliers within 20$''$ (velocity and distance limits obtained from \citet{baumgardt19}). We define the significant outliers in the velocity higher than 62 km/s based on the uncertainty in the velocity of the star.  A star would be a 1$\sigma$ outlier if its velocity is higher than 62~km/s including its uncertainty at the 1$\sigma$ level. From the histogram, we have 50 stars that have velocities~$>$62 km/s out of which 22 are 1$\sigma$ outliers, twelve are 2$\sigma$ outliers and nine are 3$\sigma$ outliers. We checked for the probability of the stars being binaries from a parallel work (Wragg et al. {\it in prep.}) and found that three out of those nine stars are likely binaries. From the six non-binary stars, we found one star that is at $\sim$20$''$ with a velocity of 104$\pm$1.5~km/s. We checked for the properties of this star on the color-magnitude diagram and it was redder than the main-sequence of the cluster, hence it is most likely a non-member of the cluster. This star also has a cluster membership probability of 0.0004 from our cluster membership analysis. 
The remaining stars are further away from the cluster center~($>$1~arcmin). Their LOS velocities range from 92--128 km/s, and out of these, one star has a cluster membership probability of 0.99 and the rest of them are close to zero. A measure of the 3-D velocities of these stars can indicate the nature of their high-velocities.
The above analysis from \citet{baumgardt19} was based on N--Body simulations that were matched to the observed sample from \citet{bellini17}. The \citet{bellini17} sample contained roughly 2900 stars within the central 20$''$. We have a similar number of stars ($\sim$2500) compared to their sample within 20$''$, but note that they treat the two velocity components of the stars as separate measurements. Hence, the number of detected high-velocity stars in our sample should be halved ($\sim$10).
Although we do not detect any significant number of high-velocity stars as predicted from these simulations, there are still some possible implications to this: 1. There is no IMBH in $\omega$~Centauri. 2. The IMBH is smaller than $\sim$5$\times$~10$^4$~M$_\odot$. 3. The existing simulations might not be fully scaled to $\omega$~Centauri. 
From \citet{baumgardt19} simulations which were for a BH of $\sim$4.7$\times$~10$^4$~M$_\odot$, the number of high-velocity stars should scale with the BH mass. Assuming that the stars within the sphere of influence (SOI) are potentially high-velocity stars, the radius of the SOI scales with the BH mass, and the volume scales with the cube of the BH mass. Therefore, reducing the BH mass by 50\% results in a decrease in the number of high-velocity stars by a factor of 8, which in our case would be 5 stars for a BH mass of 2.3$\times$10$^4$M$_\odot$. For a BH mass much lower than that, there might be no stars that we might be able to detect at all. This still does not completely rule out the no-IMBH scenario, and additional investigation using dynamical models is necessary. 

\section{Summary and Discussion}

In this paper, we used the MUSE WFM and NFM spectroscopic data to analyse the central kinematics of $\omega$~Centauri.  The next paper in the series will present dynamical models based on different centers to infer whether a central dark mass or IMBH is present.
\begin{itemize}
    \item We extracted the LOS velocities of 28,108 stars from both WFM and NFM datacubes.  The dataset is of the highest resolution yet obtained for the central 20$''$ with a spatial resolution of 25 mas.  We used the data to obtain the LOS velocity and velocity dispersion maps using KNN analysis and the LOESS method and then ran a {\it kinemetry} analysis on the maps. For the first time, our dataset revealed that the central 20$''$ of the cluster are counter-rotating with respect to the large-scale rotation that is on the scale of 100$''$, with a rotation speed of $\sim$3--4~km/s.
    \item  We used several methods to determine a kinematic center for the cluster based on the counter-rotation and the peak in the velocity dispersion and finally used the kinemetry to constrain the centers using both kinematic maps. Our center based on the counter-rotation was closest ($\sim$5$''$) to the \citetalias{vdMA10} whereas the center based on the dispersion peak was found to be near ($\sim$2.5$''$) \citetalias{N10} centers. Both centers are offset by $\sim$10$''$, which is $\sim$0.25~pc.
    \item We compared the proposed centers of $\omega$~Centauri from \citetalias{NGB08}, \citetalias{N10}, \citetalias{vdMA10}, and our kinematic center,  and used those centers to derive the radial profiles for rotations and dispersions. We found similar trends in the dispersion profiles for different centers when compared to the previously observed data.
    \item While comparing the dispersion profiles we also found that irrespective of whether the data was integrated-light kinematics or discrete velocities, the trends for the velocity dispersion profiles were similar.
    \item To search for evidence of any potential IMBH, we searched for high-velocity stars~($>$~62~km/s) close to the center~($<$~20$''$) of the cluster but did not find any significant outliers. However, this does not completely rule out the IMBH as a lower mass in IMBH can result in non-detection of any high-velocity stars.
\end{itemize}

Assuming that the center of $\omega$~Centauri contains an IMBH with a mass of 4.7$\times$10$^4$~M$_\odot$, the sphere of influence is 15$''$, which is the region of the counter-rotation. An interesting possibility for the counter-rotation is the existence of a potential binary IMBH at the center of the cluster. Simulations of three-body encounters  from \citet{mapelli05} show that a fraction of stars (55-70\%) tend to align their angular momentum to the binary IMBH. These simulations consist of three-body encounters of two BHs (binary IMBH) and a cluster star of mass 0.5$M_\odot$, where the trajectories of stars were observed that were scattered by the binary IMBH. If the binary BHs were massive enough and their orbits were wide enough and in addition, if their orbital angular momentum was higher than that of the incoming star, the star could co-rotate with its angular momentum aligned to the binary IMBH. In our case, if the binary IMBH orbital rotation is misaligned with the global rotation, there is a probability of finding stars aligned with the binary IMBH rotation and thus counter-rotating. But, this case is unlikely to explain the offset between the peak of the dispersion and the center of the counter-rotation that we find.  One scenario that can likely explain the offset in the counter-rotation and the dispersion peak is a wandering IMBH. \citet{deVita18} estimated a scaling relation between the cluster parameters and the wandering radius for an IMBH using $N$--Body models. In particular, they derived the displacements of IMBHs in the clusters using a fixed ratio for cluster mass to IMBH mass and found that they deviated on average within 1$''$. But, they also found a few outliers, specifically for $\omega$~Centauri, which had an average displacement of 2.5$''$. Note that their assumptions relied on the extrapolations of BH mass scaling relationships, where they assumed a fixed mass ratio between M$_{BH}$/M$_{tot}$ = 10$^3$. The lower mass end of the scaling relationships suffers from a significant scatter and thus the displacements can vary too. We used their Equation~16 for a BH mass of 4.7$\times$10$^4$M$_\odot$ to estimate a wandering radius $r_w$, which was $\sim$0.7$''$ for a cluster core radius of 3.6~pc and mass of 3.6$\times$10$^6$~M$_\odot$ \citep{baumgardt18}. The offset in our dispersion center and the kinematic center is $\sim$9$''$, and within 1$\sigma$ this offset drops to $\sim$3$''$. A wandering IMBH explanation is not highly unlikely although more detailed simulations tailored to $\omega$~Centauri would be needed to investigate this scenario. 

Several studies of early-type galaxies show a fraction of galaxies containing kinematically decoupled cores (KDCs) \citep[e.g.,][]{krajnovic11,cappellari16}. These are the cores of galaxies that are not aligned to the global rotation of the galaxy, which is usually a consequence of a previous merger \citep[e.g.,][]{kormendy94,hoffman10}. For example, the NSC of NGC~404 counter-rotates with respect to the galaxy\citep{seth10,nguyen17}, and this is believed to be due to a merger with a gas-rich dwarf around 1 Gyr ago. This is also supported by the young stars that are found near the center of this galaxy. Many galaxies are also observed to show this behavior such as NGC~4150, NGC~3032, NGC~4382, etc. \citep{mcdermid06}. Some GCs such as M15, which is a core-collapsed cluster \citep{crusher21}, and M53 \citep{boberg17} were also observed to have an inner rotation axis not aligned with the outer rotation similar to $\omega$~Centauri. If $\omega$~Centauri was formed through globular cluster mergers, there is a possibility for this kind of varied kinematics towards the center.  The half-mass relaxation time of $\omega$~Centauri is 9.6 Gyr \citep{harris96}, so primordial kinematic features might still be observable. However, the relaxation time close to the center will be much shorter, and thus an initial central counter-rotation should have been erased. Despite the general expectation that rotation is lost on relatively short timescales, there is considerable nuclear rotation seen in many GCs with even shorter relaxation times than $\omega$ Centauri \citep[e.g.,][]{fabricius14, kamann18}. However, in a scenario with purely isotropic mergers of GCs, the expectation would probably be like in the case of galaxy dry mergers, a velocity field consistent with zero mean velocity and thus without any surviving rotating substructures \citep[e.g.,][]{hernquist93,cox06}.  But, recently \citet{tsatsi17} found that even pure isotropic GC mergers can result in a rotating NSC, but the mechanism is not trivial. $\omega$~Centauri also contains a complex population of stars and a spread in metallicities and ages \citep[e.g.,][]{johnson10,latour21} that is suggestive of this  scenario where different populations from multiple GCs might be surviving in the present-day cluster \citep[e.g.,][]{lee94,bedin04}.

Recently, data from Gaia was used to trace the origin of $\omega$~Centauri, where \citet{majewski12,massari19,forbes20,pfeffer21} suggested it to be the former core of the $Gaia$-Enceladus/Sausage galaxy and galaxies are likely to contain central BHs.  If there were strong tidal interactions in the past, then it is a possibility that the core along with its BH could have been offset from the global rotation. If the offset of the core  is an imprint of this tidal interaction, this tidal effect must have acted along the north-south direction, e.g. when $\omega$~Centauri passed the Milky Way disk since we observe the counter-rotation elongated in that direction. These kinematics are indicative of a complex system with probable interactions in the past. However, recent studies by \citet{tiongco18,tiongco22} used $N$--Body simulations to follow the evolution of rotating star clusters in the presence of an external tidal field. They found that the dynamics within the cluster are perturbed by the tidal field, mainly a tidal torque from the host galaxy that affects the internal rotation of the cluster and introduces a precession of the cluster's rotation axis. Mostly, the inner regions are dominated by the cluster's natal rotation, which is dependent on the initial conditions, whereas precession is introduced in the outer parts of the cluster. As the cluster evolves, a radial variation of the rotation axis is observed, and depending on the initial intrinsic rotation, it can lead to a counter-rotation in the cluster. When compared, the precession would have to be observed in the outer parts of the cluster that are beyond the half-mass radius of the cluster (10.4~pc), whereas we observe the counter-rotation within the central 20$''$ (0.5~pc). 
It is highly unlikely from these simulations that the counter-rotation was caused by tidal effects. 
Our next step is to model the kinematics using Schwarzschild dynamical models. This will allow us to constrain a possible IMBH or dark remnant mass distribution in this cluster. It is challenging, however, to set up this model due to the counter-rotation, its offset from the dispersion peak, and overall, the outer kinematics may dominate the rotation field.

\section*{Acknowledgements}
R.P. and S.K acknowledge funding from UKRI in the form of a Future Leaders Fellowship (grant no. MR/T022868/1). PMW acknowledges support by the BMBF from the ErUM program
(project VLT-BlueMUSE, grant 05A20BAB)

%%%%%%%%%%%%%%%%%%%%%%%%%%%%%%%%%%%%%%%%%%%%%%%%%%
\section*{Data Availability}

The data underlying this article will be shared on reasonable request to the corresponding author.



\bibliographystyle{mnras}
\bibliography{main} % if your bibtex file is called example.bib

\begin{thebibliography}{}
\makeatletter
\relax
\def\mn@urlcharsother{\let\do\@makeother \do\$\do\&\do\#\do\^\do\_\do\%\do\~}
\def\mn@doi{\begingroup\mn@urlcharsother \@ifnextchar [ {\mn@doi@}
  {\mn@doi@[]}}
\def\mn@doi@[#1]#2{\def\@tempa{#1}\ifx\@tempa\@empty \href
  {http://dx.doi.org/#2} {doi:#2}\else \href {http://dx.doi.org/#2} {#1}\fi
  \endgroup}
\def\mn@eprint#1#2{\mn@eprint@#1:#2::\@nil}
\def\mn@eprint@arXiv#1{\href {http://arxiv.org/abs/#1} {{\tt arXiv:#1}}}
\def\mn@eprint@dblp#1{\href {http://dblp.uni-trier.de/rec/bibtex/#1.xml}
  {dblp:#1}}
\def\mn@eprint@#1:#2:#3:#4\@nil{\def\@tempa {#1}\def\@tempb {#2}\def\@tempc
  {#3}\ifx \@tempc \@empty \let \@tempc \@tempb \let \@tempb \@tempa \fi \ifx
  \@tempb \@empty \def\@tempb {arXiv}\fi \@ifundefined
  {mn@eprint@\@tempb}{\@tempb:\@tempc}{\expandafter \expandafter \csname
  mn@eprint@\@tempb\endcsname \expandafter{\@tempc}}}

\bibitem[\protect\citeauthoryear{{Ahn} et~al.,}{{Ahn} et~al.}{2017}]{ahn17}
{Ahn} C.~P.,  et~al., 2017, \mn@doi [\apj] {10.3847/1538-4357/aa6972}, \href
  {http://adsabs.harvard.edu/abs/2017ApJ...839...72A} {839, 72}

\bibitem[\protect\citeauthoryear{{Ahn} et~al.,}{{Ahn} et~al.}{2018}]{ahn18}
{Ahn} C.~P.,  et~al., 2018, \mn@doi [\apj] {10.3847/1538-4357/aabc57}, \href
  {https://ui.adsabs.harvard.edu/abs/2018ApJ...858..102A} {858, 102}

\bibitem[\protect\citeauthoryear{{Alfaro-Cuello} et~al.,}{{Alfaro-Cuello}
  et~al.}{2019}]{alfaro-cuello19}
{Alfaro-Cuello} M.,  et~al., 2019, \mn@doi [\apj] {10.3847/1538-4357/ab1b2c},
  \href {https://ui.adsabs.harvard.edu/abs/2019ApJ...886...57A} {886, 57}

\bibitem[\protect\citeauthoryear{{Anderson} \& {van der Marel}}{{Anderson} \&
  {van der Marel}}{2010}]{anderson10}
{Anderson} J.,  {van der Marel} R.~P.,  2010, \mn@doi [\apj]
  {10.1088/0004-637X/710/2/1032}, \href
  {https://ui.adsabs.harvard.edu/abs/2010ApJ...710.1032A} {710, 1032}

\bibitem[\protect\citeauthoryear{{Anderson} et~al.,}{{Anderson}
  et~al.}{2008}]{anderson08}
{Anderson} J.,  et~al., 2008, \mn@doi [\aj] {10.1088/0004-6256/135/6/2055},
  \href {https://ui.adsabs.harvard.edu/abs/2008AJ....135.2055A} {135, 2055}

\bibitem[\protect\citeauthoryear{{Baumgardt}}{{Baumgardt}}{2017}]{baumgardt17}
{Baumgardt} H.,  2017, \mn@doi [\mnras] {10.1093/mnras/stw2488}, \href
  {https://ui.adsabs.harvard.edu/abs/2017MNRAS.464.2174B} {464, 2174}

\bibitem[\protect\citeauthoryear{{Baumgardt} \& {Hilker}}{{Baumgardt} \&
  {Hilker}}{2018}]{baumgardt18}
{Baumgardt} H.,  {Hilker} M.,  2018, \mn@doi [\mnras] {10.1093/mnras/sty1057},
  \href {https://ui.adsabs.harvard.edu/abs/2018MNRAS.478.1520B} {478, 1520}

\bibitem[\protect\citeauthoryear{{Baumgardt}, {Makino}  \& {Hut}}{{Baumgardt}
  et~al.}{2005}]{baumgardt05}
{Baumgardt} H.,  {Makino} J.,   {Hut} P.,  2005, \mn@doi [\apj]
  {10.1086/426893}, \href
  {https://ui.adsabs.harvard.edu/abs/2005ApJ...620..238B} {620, 238}

\bibitem[\protect\citeauthoryear{{Baumgardt} et~al.,}{{Baumgardt}
  et~al.}{2019}]{baumgardt19}
{Baumgardt} H.,  et~al., 2019, \mn@doi [\mnras] {10.1093/mnras/stz2060}, \href
  {https://ui.adsabs.harvard.edu/abs/2019MNRAS.488.5340B} {488, 5340}

\bibitem[\protect\citeauthoryear{{Bedin}, {Piotto}, {Anderson}, {Cassisi},
  {King}, {Momany}  \& {Carraro}}{{Bedin} et~al.}{2004}]{bedin04}
{Bedin} L.~R.,  {Piotto} G.,  {Anderson} J.,  {Cassisi} S.,  {King} I.~R.,
  {Momany} Y.,   {Carraro} G.,  2004, \mn@doi [\apjl] {10.1086/420847}, \href
  {https://ui.adsabs.harvard.edu/abs/2004ApJ...605L.125B} {605, L125}

\bibitem[\protect\citeauthoryear{{Beers}, {Flynn}  \& {Gebhardt}}{{Beers}
  et~al.}{1990}]{beers90}
{Beers} T.~C.,  {Flynn} K.,   {Gebhardt} K.,  1990, \mn@doi [\aj]
  {10.1086/115487}, \href
  {https://ui.adsabs.harvard.edu/abs/1990AJ....100...32B} {100, 32}

\bibitem[\protect\citeauthoryear{{Bellini}, {Anderson}, {Bedin}, {King}, {van
  der Marel}, {Piotto}  \& {Cool}}{{Bellini} et~al.}{2017}]{bellini17}
{Bellini} A.,  {Anderson} J.,  {Bedin} L.~R.,  {King} I.~R.,  {van der Marel}
  R.~P.,  {Piotto} G.,   {Cool} A.,  2017, \mn@doi [\apj]
  {10.3847/1538-4357/aa7059}, \href
  {https://ui.adsabs.harvard.edu/abs/2017ApJ...842....6B} {842, 6}

\bibitem[\protect\citeauthoryear{{Bianchini}, {van der Marel}, {del Pino},
  {Watkins}, {Bellini}, {Fardal}, {Libralato}  \& {Sills}}{{Bianchini}
  et~al.}{2018}]{bianchini18}
{Bianchini} P.,  {van der Marel} R.~P.,  {del Pino} A.,  {Watkins} L.~L.,
  {Bellini} A.,  {Fardal} M.~A.,  {Libralato} M.,   {Sills} A.,  2018, \mn@doi
  [\mnras] {10.1093/mnras/sty2365}, \href
  {https://ui.adsabs.harvard.edu/abs/2018MNRAS.481.2125B} {481, 2125}

\bibitem[\protect\citeauthoryear{{Boberg}, {Vesperini}, {Friel}, {Tiongco}  \&
  {Varri}}{{Boberg} et~al.}{2017}]{boberg17}
{Boberg} O.~M.,  {Vesperini} E.,  {Friel} E.~D.,  {Tiongco} M.~A.,   {Varri}
  A.~L.,  2017, \mn@doi [\apj] {10.3847/1538-4357/aa7070}, \href
  {https://ui.adsabs.harvard.edu/abs/2017ApJ...841..114B} {841, 114}

\bibitem[\protect\citeauthoryear{{Bressan}, {Marigo}, {Girardi}, {Salasnich},
  {Dal Cero}, {Rubele}  \& {Nanni}}{{Bressan} et~al.}{2012}]{bressan12}
{Bressan} A.,  {Marigo} P.,  {Girardi} L.,  {Salasnich} B.,  {Dal Cero} C.,
  {Rubele} S.,   {Nanni} A.,  2012, \mn@doi [\mnras]
  {10.1111/j.1365-2966.2012.21948.x}, \href
  {http://adsabs.harvard.edu/abs/2012MNRAS.427..127B} {427, 127}

\bibitem[\protect\citeauthoryear{{Brown}, {Massey}, {Lacroix}, {Strigari},
  {Fattahi}  \& {B{\oe}hm}}{{Brown} et~al.}{2019}]{brown19}
{Brown} A.~M.,  {Massey} R.,  {Lacroix} T.,  {Strigari} L.~E.,  {Fattahi} A.,
  {B{\oe}hm} C.,  2019, \mn@doi [arXiv e-prints] {10.48550/arXiv.1907.08564},
  \href {https://ui.adsabs.harvard.edu/abs/2019arXiv190708564B} {p.
  arXiv:1907.08564}

\bibitem[\protect\citeauthoryear{{Cappellari}}{{Cappellari}}{2016}]{cappellari16}
{Cappellari} M.,  2016, \mn@doi [\araa] {10.1146/annurev-astro-082214-122432},
  \href {https://ui.adsabs.harvard.edu/abs/2016ARA&A..54..597C} {54, 597}

\bibitem[\protect\citeauthoryear{{Cappellari} et~al.,}{{Cappellari}
  et~al.}{2013}]{atlas3d20}
{Cappellari} M.,  et~al., 2013, \mn@doi [\mnras] {10.1093/mnras/stt644}, \href
  {http://adsabs.harvard.edu/abs/2013MNRAS.432.1862C} {432, 1862}

\bibitem[\protect\citeauthoryear{Cleveland}{Cleveland}{1979}]{cleveland79}
Cleveland W.~S.,  1979, \mn@doi [Journal of the American Statistical
  Association] {10.1080/01621459.1979.10481038}, 74, 829

\bibitem[\protect\citeauthoryear{Cleveland \& Devlin}{Cleveland \&
  Devlin}{1988}]{cleveland88}
Cleveland W.~S.,  Devlin S.~J.,  1988, \mn@doi [Journal of the American
  Statistical Association] {10.1080/01621459.1988.10478639}, 83, 596

\bibitem[\protect\citeauthoryear{{Cox}, {Dutta}, {Di Matteo}, {Hernquist},
  {Hopkins}, {Robertson}  \& {Springel}}{{Cox} et~al.}{2006}]{cox06}
{Cox} T.~J.,  {Dutta} S.~N.,  {Di Matteo} T.,  {Hernquist} L.,  {Hopkins}
  P.~F.,  {Robertson} B.,   {Springel} V.,  2006, \mn@doi [\apj]
  {10.1086/507474}, \href
  {https://ui.adsabs.harvard.edu/abs/2006ApJ...650..791C} {650, 791}

\bibitem[\protect\citeauthoryear{{Evans}, {Strigari}  \& {Zivick}}{{Evans}
  et~al.}{2022}]{evans22}
{Evans} A.~J.,  {Strigari} L.~E.,   {Zivick} P.,  2022, \mn@doi [\mnras]
  {10.1093/mnras/stac261}, \href
  {https://ui.adsabs.harvard.edu/abs/2022MNRAS.511.4251E} {511, 4251}

\bibitem[\protect\citeauthoryear{{Fabricius} et~al.,}{{Fabricius}
  et~al.}{2014}]{fabricius14}
{Fabricius} M.~H.,  et~al., 2014, \mn@doi [\apjl]
  {10.1088/2041-8205/787/2/L26}, \href
  {https://ui.adsabs.harvard.edu/abs/2014ApJ...787L..26F} {787, L26}

\bibitem[\protect\citeauthoryear{{F{\'e}tick} et~al.,}{{F{\'e}tick}
  et~al.}{2019}]{maoppy}
{F{\'e}tick} R.~J.~L.,  et~al., 2019, \mn@doi [\aap]
  {10.1051/0004-6361/201935830}, \href
  {https://ui.adsabs.harvard.edu/abs/2019A&A...628A..99F} {628, A99}

\bibitem[\protect\citeauthoryear{{Fiestas}, {Spurzem}  \& {Kim}}{{Fiestas}
  et~al.}{2006}]{fiestas06}
{Fiestas} J.,  {Spurzem} R.,   {Kim} E.,  2006, \mn@doi [\mnras]
  {10.1111/j.1365-2966.2006.11036.x}, \href
  {https://ui.adsabs.harvard.edu/abs/2006MNRAS.373..677F} {373, 677}

\bibitem[\protect\citeauthoryear{{Forbes}}{{Forbes}}{2020}]{forbes20}
{Forbes} D.~A.,  2020, \mn@doi [\mnras] {10.1093/mnras/staa245}, \href
  {https://ui.adsabs.harvard.edu/abs/2020MNRAS.493..847F} {493, 847}

\bibitem[\protect\citeauthoryear{Foreman-Mackey, Hogg, Lang  \&
  Goodman}{Foreman-Mackey et~al.}{2013}]{foreman13}
Foreman-Mackey D.,  Hogg D.~W.,  Lang D.,   Goodman J.,  2013, \mn@doi
  [Publications of the Astronomical Society of the Pacific] {10.1086/670067},
  125, 306–312

\bibitem[\protect\citeauthoryear{{Freeman} \& {Rodgers}}{{Freeman} \&
  {Rodgers}}{1975}]{freeman75}
{Freeman} K.~C.,  {Rodgers} A.~W.,  1975, \mn@doi [\apjl] {10.1086/181945},
  \href {https://ui.adsabs.harvard.edu/abs/1975ApJ...201L..71F} {201, L71}

\bibitem[\protect\citeauthoryear{{Fusco} et~al.,}{{Fusco}
  et~al.}{2020}]{fusco20}
{Fusco} T.,  et~al., 2020, \mn@doi [\aap] {10.1051/0004-6361/202037595}, \href
  {https://ui.adsabs.harvard.edu/abs/2020A&A...635A.208F} {635, A208}

\bibitem[\protect\citeauthoryear{{Gebhardt}, {Rich}  \& {Ho}}{{Gebhardt}
  et~al.}{2005}]{gebhardt05}
{Gebhardt} K.,  {Rich} R.~M.,   {Ho} L.~C.,  2005, \mn@doi [\apj]
  {10.1086/497023}, \href
  {https://ui.adsabs.harvard.edu/abs/2005ApJ...634.1093G} {634, 1093}

\bibitem[\protect\citeauthoryear{{Geyer}, {Hopp}  \& {Nelles}}{{Geyer}
  et~al.}{1983}]{geyer83}
{Geyer} E.~H.,  {Hopp} U.,   {Nelles} B.,  1983, \aap, \href
  {https://ui.adsabs.harvard.edu/abs/1983A&A...125..359G} {125, 359}

\bibitem[\protect\citeauthoryear{{Giesers} et~al.,}{{Giesers}
  et~al.}{2019}]{giesers19}
{Giesers} B.,  et~al., 2019, \mn@doi [\aap] {10.1051/0004-6361/201936203},
  \href {https://ui.adsabs.harvard.edu/abs/2019A&A...632A...3G} {632, A3}

\bibitem[\protect\citeauthoryear{{G{\"o}ttgens} et~al.,}{{G{\"o}ttgens}
  et~al.}{2021}]{goettgens21}
{G{\"o}ttgens} F.,  et~al., 2021, \mn@doi [\mnras] {10.1093/mnras/stab2449},
  \href {https://ui.adsabs.harvard.edu/abs/2021MNRAS.507.4788G} {507, 4788}

\bibitem[\protect\citeauthoryear{{Gratton}, {Carretta}  \&
  {Bragaglia}}{{Gratton} et~al.}{2012}]{gratton12}
{Gratton} R.~G.,  {Carretta} E.,   {Bragaglia} A.,  2012, \mn@doi [\aapr]
  {10.1007/s00159-012-0050-3}, \href
  {https://ui.adsabs.harvard.edu/abs/2012A&ARv..20...50G} {20, 50}

\bibitem[\protect\citeauthoryear{{Greene}, {Strader}  \& {Ho}}{{Greene}
  et~al.}{2020}]{greene20}
{Greene} J.~E.,  {Strader} J.,   {Ho} L.~C.,  2020, \mn@doi [\araa]
  {10.1146/annurev-astro-032620-021835}, \href
  {https://ui.adsabs.harvard.edu/abs/2020ARA&A..58..257G} {58, 257}

\bibitem[\protect\citeauthoryear{{Habouzit} et~al.,}{{Habouzit}
  et~al.}{2021}]{habouzit21}
{Habouzit} M.,  et~al., 2021, \mn@doi [\mnras] {10.1093/mnras/stab496}, \href
  {https://ui.adsabs.harvard.edu/abs/2021MNRAS.503.1940H} {503, 1940}

\bibitem[\protect\citeauthoryear{{Harris}}{{Harris}}{1996}]{harris96}
{Harris} W.~E.,  1996, \mn@doi [\aj] {10.1086/118116}, \href
  {https://ui.adsabs.harvard.edu/abs/1996AJ....112.1487H} {112, 1487}

\bibitem[\protect\citeauthoryear{{Helmi}, {Babusiaux}, {Koppelman}, {Massari},
  {Veljanoski}  \& {Brown}}{{Helmi} et~al.}{2018}]{helmi18}
{Helmi} A.,  {Babusiaux} C.,  {Koppelman} H.~H.,  {Massari} D.,  {Veljanoski}
  J.,   {Brown} A. G.~A.,  2018, \mn@doi [\nat] {10.1038/s41586-018-0625-x},
  \href {https://ui.adsabs.harvard.edu/abs/2018Natur.563...85H} {563, 85}

\bibitem[\protect\citeauthoryear{{Hernquist}}{{Hernquist}}{1993}]{hernquist93}
{Hernquist} L.,  1993, \mn@doi [\apj] {10.1086/172686}, \href
  {https://ui.adsabs.harvard.edu/abs/1993ApJ...409..548H} {409, 548}

\bibitem[\protect\citeauthoryear{{Hoffman}, {Cox}, {Dutta}  \&
  {Hernquist}}{{Hoffman} et~al.}{2010}]{hoffman10}
{Hoffman} L.,  {Cox} T.~J.,  {Dutta} S.,   {Hernquist} L.,  2010, \mn@doi
  [\apj] {10.1088/0004-637X/723/1/818}, \href
  {https://ui.adsabs.harvard.edu/abs/2010ApJ...723..818H} {723, 818}

\bibitem[\protect\citeauthoryear{{Husser}, {Wende-von Berg}, {Dreizler},
  {Homeier}, {Reiners}, {Barman}  \& {Hauschildt}}{{Husser}
  et~al.}{2013}]{husser13}
{Husser} T.~O.,  {Wende-von Berg} S.,  {Dreizler} S.,  {Homeier} D.,  {Reiners}
  A.,  {Barman} T.,   {Hauschildt} P.~H.,  2013, \mn@doi [\aap]
  {10.1051/0004-6361/201219058}, \href
  {https://ui.adsabs.harvard.edu/abs/2013A&A...553A...6H} {553, A6}

\bibitem[\protect\citeauthoryear{{Husser} et~al.,}{{Husser}
  et~al.}{2016}]{husser16}
{Husser} T.-O.,  et~al., 2016, \mn@doi [\aap] {10.1051/0004-6361/201526949},
  \href {https://ui.adsabs.harvard.edu/abs/2016A&A...588A.148H} {588, A148}

\bibitem[\protect\citeauthoryear{{Husser} et~al.,}{{Husser}
  et~al.}{2020}]{husser20}
{Husser} T.-O.,  et~al., 2020, \mn@doi [\aap] {10.1051/0004-6361/201936508},
  \href {https://ui.adsabs.harvard.edu/abs/2020A&A...635A.114H} {635, A114}

\bibitem[\protect\citeauthoryear{{Ibata} et~al.,}{{Ibata}
  et~al.}{2009}]{ibata09}
{Ibata} R.,  et~al., 2009, \mn@doi [\apjl] {10.1088/0004-637X/699/2/L169},
  \href {https://ui.adsabs.harvard.edu/abs/2009ApJ...699L.169I} {699, L169}

\bibitem[\protect\citeauthoryear{{Jedrzejewski}}{{Jedrzejewski}}{1987}]{Jedrzejewski87}
{Jedrzejewski} R.~I.,  1987, \mn@doi [\mnras] {10.1093/mnras/226.4.747}, \href
  {https://ui.adsabs.harvard.edu/abs/1987MNRAS.226..747J} {226, 747}

\bibitem[\protect\citeauthoryear{{Johnson} \& {Pilachowski}}{{Johnson} \&
  {Pilachowski}}{2010a}]{j&p10}
{Johnson} C.~I.,  {Pilachowski} C.~A.,  2010a, \mn@doi [\apj]
  {10.1088/0004-637X/722/2/1373}, \href
  {https://ui.adsabs.harvard.edu/abs/2010ApJ...722.1373J} {722, 1373}

\bibitem[\protect\citeauthoryear{{Johnson} \& {Pilachowski}}{{Johnson} \&
  {Pilachowski}}{2010b}]{johnson10}
{Johnson} C.~I.,  {Pilachowski} C.~A.,  2010b, \mn@doi [\apj]
  {10.1088/0004-637X/722/2/1373}, \href
  {https://ui.adsabs.harvard.edu/abs/2010ApJ...722.1373J} {722, 1373}

\bibitem[\protect\citeauthoryear{{Kamann}}{{Kamann}}{2018}]{pampelmuse}
{Kamann} S.,  2018, {PampelMuse: Crowded-field 3D spectroscopy} (\mn@eprint
  {ascl} {1805.021})

\bibitem[\protect\citeauthoryear{{Kamann}, {Wisotzki}  \& {Roth}}{{Kamann}
  et~al.}{2013}]{kamann13}
{Kamann} S.,  {Wisotzki} L.,   {Roth} M.~M.,  2013, \mn@doi [\aap]
  {10.1051/0004-6361/201220476}, \href
  {https://ui.adsabs.harvard.edu/abs/2013A&A...549A..71K} {549, A71}

\bibitem[\protect\citeauthoryear{{Kamann} et~al.,}{{Kamann}
  et~al.}{2016}]{kamann16}
{Kamann} S.,  et~al., 2016, \mn@doi [\aap] {10.1051/0004-6361/201527065}, \href
  {https://ui.adsabs.harvard.edu/abs/2016A&A...588A.149K} {588, A149}

\bibitem[\protect\citeauthoryear{{Kamann} et~al.,}{{Kamann}
  et~al.}{2018}]{kamann18}
{Kamann} S.,  et~al., 2018, \mn@doi [\mnras] {10.1093/mnras/stx2719}, \href
  {https://ui.adsabs.harvard.edu/abs/2018MNRAS.473.5591K} {473, 5591}

\bibitem[\protect\citeauthoryear{{Kelly}}{{Kelly}}{2007}]{kelly07}
{Kelly} B.~C.,  2007, \mn@doi [\apj] {10.1086/519947}, \href
  {https://ui.adsabs.harvard.edu/abs/2007ApJ...665.1489K} {665, 1489}

\bibitem[\protect\citeauthoryear{{Kormendy}, {Dressler}, {Byun}, {Faber},
  {Grillmair}, {Lauer}, {Richstone}  \& {Tremaine}}{{Kormendy}
  et~al.}{1994}]{kormendy94}
{Kormendy} J.,  {Dressler} A.,  {Byun} Y.~I.,  {Faber} S.~M.,  {Grillmair} C.,
  {Lauer} T.~R.,  {Richstone} D.,   {Tremaine} S.,  1994, in European Southern
  Observatory Conference and Workshop Proceedings. p.~147

\bibitem[\protect\citeauthoryear{{Krajnovi{\'c}}, {Cappellari}, {de Zeeuw}  \&
  {Copin}}{{Krajnovi{\'c}} et~al.}{2006}]{krajnovic06}
{Krajnovi{\'c}} D.,  {Cappellari} M.,  {de Zeeuw} P.~T.,   {Copin} Y.,  2006,
  \mn@doi [\mnras] {10.1111/j.1365-2966.2005.09902.x}, \href
  {https://ui.adsabs.harvard.edu/abs/2006MNRAS.366..787K} {366, 787}

\bibitem[\protect\citeauthoryear{{Krajnovi{\'c}} et~al.,}{{Krajnovi{\'c}}
  et~al.}{2008}]{krajnovic08}
{Krajnovi{\'c}} D.,  et~al., 2008, \mn@doi [\mnras]
  {10.1111/j.1365-2966.2008.13712.x}, \href
  {https://ui.adsabs.harvard.edu/abs/2008MNRAS.390...93K} {390, 93}

\bibitem[\protect\citeauthoryear{{Krajnovi{\'c}} et~al.,}{{Krajnovi{\'c}}
  et~al.}{2011}]{krajnovic11}
{Krajnovi{\'c}} D.,  et~al., 2011, \mn@doi [\mnras]
  {10.1111/j.1365-2966.2011.18560.x}, \href
  {https://ui.adsabs.harvard.edu/abs/2011MNRAS.414.2923K} {414, 2923}

\bibitem[\protect\citeauthoryear{{Kruijssen} \& {Portegies Zwart}}{{Kruijssen}
  \& {Portegies Zwart}}{2009}]{kruijssen09}
{Kruijssen} J.~M.~D.,  {Portegies Zwart} S.~F.,  2009, \mn@doi [\apjl]
  {10.1088/0004-637X/698/2/L158}, \href
  {https://ui.adsabs.harvard.edu/abs/2009ApJ...698L.158K} {698, L158}

\bibitem[\protect\citeauthoryear{{Lanzoni} et~al.,}{{Lanzoni}
  et~al.}{2013}]{lanzoni13}
{Lanzoni} B.,  et~al., 2013, \mn@doi [\apj] {10.1088/0004-637X/769/2/107},
  \href {https://ui.adsabs.harvard.edu/abs/2013ApJ...769..107L} {769, 107}

\bibitem[\protect\citeauthoryear{{Latour}, {Calamida}, {Husser}, {Kamann},
  {Dreizler}  \& {Brinchmann}}{{Latour} et~al.}{2021}]{latour21}
{Latour} M.,  {Calamida} A.,  {Husser} T.~O.,  {Kamann} S.,  {Dreizler} S.,
  {Brinchmann} J.,  2021, \mn@doi [\aap] {10.1051/0004-6361/202141791}, \href
  {https://ui.adsabs.harvard.edu/abs/2021A&A...653L...8L} {653, L8}

\bibitem[\protect\citeauthoryear{{Leaman}}{{Leaman}}{2012}]{leaman12}
{Leaman} R.,  2012, \mn@doi [\aj] {10.1088/0004-6256/144/6/183}, \href
  {https://ui.adsabs.harvard.edu/abs/2012AJ....144..183L} {144, 183}

\bibitem[\protect\citeauthoryear{{Lee}, {Joo}, {Sohn}, {Rey}, {Lee}  \&
  {Walker}}{{Lee} et~al.}{1999}]{lee94}
{Lee} Y.~W.,  {Joo} J.~M.,  {Sohn} Y.~J.,  {Rey} S.~C.,  {Lee} H.~C.,
  {Walker} A.~R.,  1999, \mn@doi [\nat] {10.1038/46985}, \href
  {https://ui.adsabs.harvard.edu/abs/1999Natur.402...55L} {402, 55}

\bibitem[\protect\citeauthoryear{{Lee}, {Rey}, {Ree}, {Joo}, {Sohn}  \&
  {Yoon}}{{Lee} et~al.}{2002}]{lee02}
{Lee} Y.~W.,  {Rey} S.~C.,  {Ree} C.~H.,  {Joo} J.~M.,  {Sohn} Y.~J.,   {Yoon}
  S.~J.,  2002, in {van Leeuwen} F.,  {Hughes} J.~D.,   {Piotto} G.,  eds,
  Astronomical Society of the Pacific Conference Series Vol. 265, Omega
  Centauri, A Unique Window into Astrophysics. p.~305 (\mn@eprint {arXiv}
  {astro-ph/0110688})

\bibitem[\protect\citeauthoryear{{L{\"u}tzgendorf}, {Kissler-Patig}, {Noyola},
  {Jalali}, {de Zeeuw}, {Gebhardt}  \& {Baumgardt}}{{L{\"u}tzgendorf}
  et~al.}{2011}]{lutzgendorf11}
{L{\"u}tzgendorf} N.,  {Kissler-Patig} M.,  {Noyola} E.,  {Jalali} B.,  {de
  Zeeuw} P.~T.,  {Gebhardt} K.,   {Baumgardt} H.,  2011, \mn@doi [\aap]
  {10.1051/0004-6361/201116618}, \href
  {https://ui.adsabs.harvard.edu/abs/2011A&A...533A..36L} {533, A36}

\bibitem[\protect\citeauthoryear{{L{\"u}tzgendorf}, {Gebhardt}, {Baumgardt},
  {Noyola}, {Neumayer}, {Kissler-Patig}  \& {de Zeeuw}}{{L{\"u}tzgendorf}
  et~al.}{2015}]{luetzgendorf15}
{L{\"u}tzgendorf} N.,  {Gebhardt} K.,  {Baumgardt} H.,  {Noyola} E.,
  {Neumayer} N.,  {Kissler-Patig} M.,   {de Zeeuw} T.,  2015, \mn@doi [\aap]
  {10.1051/0004-6361/201425524}, \href
  {https://ui.adsabs.harvard.edu/abs/2015A&A...581A...1L} {581, A1}

\bibitem[\protect\citeauthoryear{{Lynden-Bell}}{{Lynden-Bell}}{1967}]{lynden-bell67}
{Lynden-Bell} D.,  1967, \mn@doi [\mnras] {10.1093/mnras/136.1.101}, \href
  {https://ui.adsabs.harvard.edu/abs/1967MNRAS.136..101L} {136, 101}

\bibitem[\protect\citeauthoryear{{Majewski}, {Nidever}, {Smith}, {Damke},
  {Kunkel}, {Patterson}, {Bizyaev}  \& {Garc{\'\i}a P{\'e}rez}}{{Majewski}
  et~al.}{2012}]{majewski12}
{Majewski} S.~R.,  {Nidever} D.~L.,  {Smith} V.~V.,  {Damke} G.~J.,  {Kunkel}
  W.~E.,  {Patterson} R.~J.,  {Bizyaev} D.,   {Garc{\'\i}a P{\'e}rez} A.~E.,
  2012, \mn@doi [\apjl] {10.1088/2041-8205/747/2/L37}, \href
  {https://ui.adsabs.harvard.edu/abs/2012ApJ...747L..37M} {747, L37}

\bibitem[\protect\citeauthoryear{{Mapelli}, {Colpi}, {Possenti}  \&
  {Sigurdsson}}{{Mapelli} et~al.}{2005}]{mapelli05}
{Mapelli} M.,  {Colpi} M.,  {Possenti} A.,   {Sigurdsson} S.,  2005, \mn@doi
  [\mnras] {10.1111/j.1365-2966.2005.09653.x}, \href
  {https://ui.adsabs.harvard.edu/abs/2005MNRAS.364.1315M} {364, 1315}

\bibitem[\protect\citeauthoryear{{Martocchia} et~al.,}{{Martocchia}
  et~al.}{2018}]{martocchia18}
{Martocchia} S.,  et~al., 2018, \mn@doi [\mnras] {10.1093/mnras/stx2556}, \href
  {https://ui.adsabs.harvard.edu/abs/2018MNRAS.473.2688M} {473, 2688}

\bibitem[\protect\citeauthoryear{{Massari}, {Koppelman}  \& {Helmi}}{{Massari}
  et~al.}{2019}]{massari19}
{Massari} D.,  {Koppelman} H.~H.,   {Helmi} A.,  2019, \mn@doi [\aap]
  {10.1051/0004-6361/201936135}, \href
  {https://ui.adsabs.harvard.edu/abs/2019A&A...630L...4M} {630, L4}

\bibitem[\protect\citeauthoryear{{Mayor} et~al.,}{{Mayor}
  et~al.}{1997}]{mayor97}
{Mayor} M.,  et~al., 1997, \mn@doi [\aj] {10.1086/118539}, \href
  {https://ui.adsabs.harvard.edu/abs/1997AJ....114.1087M} {114, 1087}

\bibitem[\protect\citeauthoryear{{McDermid} et~al.,}{{McDermid}
  et~al.}{2006}]{mcdermid06}
{McDermid} R.~M.,  et~al., 2006, \mn@doi [\mnras]
  {10.1111/j.1365-2966.2006.11065.x}, \href
  {https://ui.adsabs.harvard.edu/abs/2006MNRAS.373..906M} {373, 906}

\bibitem[\protect\citeauthoryear{{Merritt}, {Meylan}  \& {Mayor}}{{Merritt}
  et~al.}{1997}]{merritt97}
{Merritt} D.,  {Meylan} G.,   {Mayor} M.,  1997, \mn@doi [\aj]
  {10.1086/118538}, \href
  {https://ui.adsabs.harvard.edu/abs/1997AJ....114.1074M} {114, 1074}

\bibitem[\protect\citeauthoryear{{Meylan} \& {Mayor}}{{Meylan} \&
  {Mayor}}{1986}]{mm86}
{Meylan} G.,  {Mayor} M.,  1986, \aap, \href
  {https://ui.adsabs.harvard.edu/abs/1986A&A...166..122M} {166, 122}

\bibitem[\protect\citeauthoryear{{Meylan}, {Mayor}, {Duquennoy}  \&
  {Dubath}}{{Meylan} et~al.}{1995}]{meylan95}
{Meylan} G.,  {Mayor} M.,  {Duquennoy} A.,   {Dubath} P.,  1995, \aap, \href
  {https://ui.adsabs.harvard.edu/abs/1995A&A...303..761M} {303, 761}

\bibitem[\protect\citeauthoryear{{Nguyen} et~al.,}{{Nguyen}
  et~al.}{2017a}]{ngyuen17}
{Nguyen} D.~D.,  et~al., 2017a, \mn@doi [\apj] {10.3847/1538-4357/aa5cb4},
  \href {http://adsabs.harvard.edu/abs/2017ApJ...836..237N} {836, 237}

\bibitem[\protect\citeauthoryear{{Nguyen} et~al.,}{{Nguyen}
  et~al.}{2017b}]{nguyen17}
{Nguyen} D.~D.,  et~al., 2017b, \mn@doi [\apj] {10.3847/1538-4357/aa5cb4},
  \href {https://ui.adsabs.harvard.edu/abs/2017ApJ...836..237N} {836, 237}

\bibitem[\protect\citeauthoryear{{Nguyen} et~al.,}{{Nguyen}
  et~al.}{2018}]{nguyen18}
{Nguyen} D.~D.,  et~al., 2018, \mn@doi [\apj] {10.3847/1538-4357/aabe28}, \href
  {http://adsabs.harvard.edu/abs/2018ApJ...858..118N} {858, 118}

\bibitem[\protect\citeauthoryear{{Nguyen} et~al.,}{{Nguyen}
  et~al.}{2019}]{nguyen19}
{Nguyen} D.~D.,  et~al., 2019, \mn@doi [\apj] {10.3847/1538-4357/aafe7a}, \href
  {https://ui.adsabs.harvard.edu/abs/2019ApJ...872..104N} {872, 104}

\bibitem[\protect\citeauthoryear{Nitschai et~al.,}{Nitschai
  et~al.}{2023}]{nitschai23}
Nitschai M.~S.,  et~al., 2023, oMEGACat I: MUSE spectroscopy of 300,000 stars
  within the half-light radius of $\omega$ Centauri (\mn@eprint {arXiv}
  {2309.02503})

\bibitem[\protect\citeauthoryear{{Noyola}, {Gebhardt}  \& {Bergmann}}{{Noyola}
  et~al.}{2008}]{NGB08}
{Noyola} E.,  {Gebhardt} K.,   {Bergmann} M.,  2008, \mn@doi [\apj]
  {10.1086/529002}, \href
  {https://ui.adsabs.harvard.edu/abs/2008ApJ...676.1008N} {676, 1008}

\bibitem[\protect\citeauthoryear{{Noyola}, {Gebhardt}, {Kissler-Patig},
  {L{\"u}tzgendorf}, {Jalali}, {de Zeeuw}  \& {Baumgardt}}{{Noyola}
  et~al.}{2010}]{N10}
{Noyola} E.,  {Gebhardt} K.,  {Kissler-Patig} M.,  {L{\"u}tzgendorf} N.,
  {Jalali} B.,  {de Zeeuw} P.~T.,   {Baumgardt} H.,  2010, \mn@doi [\apjl]
  {10.1088/2041-8205/719/1/L60}, \href
  {https://ui.adsabs.harvard.edu/abs/2010ApJ...719L..60N} {719, L60}

\bibitem[\protect\citeauthoryear{{Pechetti} et~al.,}{{Pechetti}
  et~al.}{2022}]{pechetti22}
{Pechetti} R.,  et~al., 2022, \mn@doi [\apj] {10.3847/1538-4357/ac339f}, \href
  {https://ui.adsabs.harvard.edu/abs/2022ApJ...924...48P} {924, 48}

\bibitem[\protect\citeauthoryear{Pedregosa et~al.,}{Pedregosa
  et~al.}{2011}]{pedregosa11}
Pedregosa F.,  et~al., 2011, Journal of Machine Learning Research, 12, 2825

\bibitem[\protect\citeauthoryear{{Pfeffer}, {Lardo}, {Bastian}, {Saracino}  \&
  {Kamann}}{{Pfeffer} et~al.}{2021}]{pfeffer21}
{Pfeffer} J.,  {Lardo} C.,  {Bastian} N.,  {Saracino} S.,   {Kamann} S.,  2021,
  \mn@doi [\mnras] {10.1093/mnras/staa3407}, \href
  {https://ui.adsabs.harvard.edu/abs/2021MNRAS.500.2514P} {500, 2514}

\bibitem[\protect\citeauthoryear{{Piotto} et~al.,}{{Piotto}
  et~al.}{2015}]{piotto15}
{Piotto} G.,  et~al., 2015, \mn@doi [\aj] {10.1088/0004-6256/149/3/91}, \href
  {https://ui.adsabs.harvard.edu/abs/2015AJ....149...91P} {149, 91}

\bibitem[\protect\citeauthoryear{{Pryor} \& {Meylan}}{{Pryor} \&
  {Meylan}}{1993}]{pryor93}
{Pryor} C.,  {Meylan} G.,  1993, in {Djorgovski} S.~G.,  {Meylan} G.,  eds,
  Astronomical Society of the Pacific Conference Series Vol. 50, Structure and
  Dynamics of Globular Clusters. p.~357

\bibitem[\protect\citeauthoryear{{Reijns}, {Seitzer}, {Arnold}, {Freeman},
  {Ingerson}, {van den Bosch}, {van de Ven}  \& {de Zeeuw}}{{Reijns}
  et~al.}{2006}]{reijns06}
{Reijns} R.~A.,  {Seitzer} P.,  {Arnold} R.,  {Freeman} K.~C.,  {Ingerson} T.,
  {van den Bosch} R.~C.~E.,  {van de Ven} G.,   {de Zeeuw} P.~T.,  2006,
  \mn@doi [\aap] {10.1051/0004-6361:20053059}, \href
  {https://ui.adsabs.harvard.edu/abs/2006A&A...445..503R} {445, 503}

\bibitem[\protect\citeauthoryear{{Robin}, {Reyl{\'e}}, {Derri{\`e}re}  \&
  {Picaud}}{{Robin} et~al.}{2003}]{robin03}
{Robin} A.~C.,  {Reyl{\'e}} C.,  {Derri{\`e}re} S.,   {Picaud} S.,  2003,
  \mn@doi [\aap] {10.1051/0004-6361:20031117}, \href
  {https://ui.adsabs.harvard.edu/abs/2003A&A...409..523R} {409, 523}

\bibitem[\protect\citeauthoryear{{Saglia} et~al.,}{{Saglia}
  et~al.}{2016}]{saglia16}
{Saglia} R.~P.,  et~al., 2016, \mn@doi [\apj] {10.3847/0004-637X/818/1/47},
  \href {http://adsabs.harvard.edu/abs/2016ApJ...818...47S} {818, 47}

\bibitem[\protect\citeauthoryear{{Sarajedini} et~al.,}{{Sarajedini}
  et~al.}{2007}]{sarajedini07}
{Sarajedini} A.,  et~al., 2007, \mn@doi [\aj] {10.1086/511979}, \href
  {https://ui.adsabs.harvard.edu/abs/2007AJ....133.1658S} {133, 1658}

\bibitem[\protect\citeauthoryear{{Seth} et~al.,}{{Seth} et~al.}{2010}]{seth10}
{Seth} A.~C.,  et~al., 2010, \mn@doi [\apj] {10.1088/0004-637X/714/1/713},
  \href {http://adsabs.harvard.edu/abs/2010ApJ...714..713S} {714, 713}

\bibitem[\protect\citeauthoryear{{Seth} et~al.,}{{Seth} et~al.}{2014}]{seth14}
{Seth} A.~C.,  et~al., 2014, \mn@doi [\nat] {10.1038/nature13762}, \href
  {http://adsabs.harvard.edu/abs/2014Natur.513..398S} {513, 398}

\bibitem[\protect\citeauthoryear{{Sollima}, {Baumgardt}  \& {Hilker}}{{Sollima}
  et~al.}{2019}]{sollima19}
{Sollima} A.,  {Baumgardt} H.,   {Hilker} M.,  2019, \mn@doi [\mnras]
  {10.1093/mnras/stz505}, \href
  {https://ui.adsabs.harvard.edu/abs/2019MNRAS.485.1460S} {485, 1460}

\bibitem[\protect\citeauthoryear{{Suntzeff} \& {Kraft}}{{Suntzeff} \&
  {Kraft}}{1996}]{suntzeff96}
{Suntzeff} N.~B.,  {Kraft} R.~P.,  1996, \mn@doi [\aj] {10.1086/117930}, \href
  {https://ui.adsabs.harvard.edu/abs/1996AJ....111.1913S} {111, 1913}

\bibitem[\protect\citeauthoryear{{Tiongco}, {Vesperini}  \& {Varri}}{{Tiongco}
  et~al.}{2018}]{tiongco18}
{Tiongco} M.~A.,  {Vesperini} E.,   {Varri} A.~L.,  2018, \mn@doi [\mnras]
  {10.1093/mnrasl/sly009}, \href
  {https://ui.adsabs.harvard.edu/abs/2018MNRAS.475L..86T} {475, L86}

\bibitem[\protect\citeauthoryear{{Tiongco}, {Vesperini}  \& {Varri}}{{Tiongco}
  et~al.}{2022}]{tiongco22}
{Tiongco} M.~A.,  {Vesperini} E.,   {Varri} A.~L.,  2022, \mn@doi [\mnras]
  {10.1093/mnras/stac643}, \href
  {https://ui.adsabs.harvard.edu/abs/2022MNRAS.512.1584T} {512, 1584}

\bibitem[\protect\citeauthoryear{{Tsatsi}, {Mastrobuono-Battisti}, {van de
  Ven}, {Perets}, {Bianchini}  \& {Neumayer}}{{Tsatsi} et~al.}{2017}]{tsatsi17}
{Tsatsi} A.,  {Mastrobuono-Battisti} A.,  {van de Ven} G.,  {Perets} H.~B.,
  {Bianchini} P.,   {Neumayer} N.,  2017, \mn@doi [\mnras]
  {10.1093/mnras/stw2593}, \href
  {https://ui.adsabs.harvard.edu/abs/2017MNRAS.464.3720T} {464, 3720}

\bibitem[\protect\citeauthoryear{{Usher}, {Kamann}, {Gieles},
  {H{\'e}nault-Brunet}, {Dalessandro}, {Balbinot}  \& {Sollima}}{{Usher}
  et~al.}{2021}]{crusher21}
{Usher} C.,  {Kamann} S.,  {Gieles} M.,  {H{\'e}nault-Brunet} V.,
  {Dalessandro} E.,  {Balbinot} E.,   {Sollima} A.,  2021, \mn@doi [\mnras]
  {10.1093/mnras/stab565}, \href
  {https://ui.adsabs.harvard.edu/abs/2021MNRAS.503.1680U} {503, 1680}

\bibitem[\protect\citeauthoryear{{Volonteri}}{{Volonteri}}{2010}]{volonteri10}
{Volonteri} M.,  2010, \mn@doi [\aapr] {10.1007/s00159-010-0029-x}, \href
  {https://ui.adsabs.harvard.edu/abs/2010A&ARv..18..279V} {18, 279}

\bibitem[\protect\citeauthoryear{{Walker}, {Mateo}, {Olszewski}, {Sen}  \&
  {Woodroofe}}{{Walker} et~al.}{2009}]{walker09}
{Walker} M.~G.,  {Mateo} M.,  {Olszewski} E.~W.,  {Sen} B.,   {Woodroofe} M.,
  2009, \mn@doi [\aj] {10.1088/0004-6256/137/2/3109}, \href
  {https://ui.adsabs.harvard.edu/abs/2009AJ....137.3109W} {137, 3109}

\bibitem[\protect\citeauthoryear{{Weilbacher} et~al.,}{{Weilbacher}
  et~al.}{2020}]{weilbacher2020}
{Weilbacher} P.~M.,  et~al., 2020, \mn@doi [\aap]
  {10.1051/0004-6361/202037855}, \href
  {https://ui.adsabs.harvard.edu/abs/2020A&A...641A..28W} {641, A28}

\bibitem[\protect\citeauthoryear{{Zocchi}, {Gieles}  \&
  {H{\'e}nault-Brunet}}{{Zocchi} et~al.}{2019}]{zocchi19}
{Zocchi} A.,  {Gieles} M.,   {H{\'e}nault-Brunet} V.,  2019, \mn@doi [\mnras]
  {10.1093/mnras/sty1508}, \href
  {https://ui.adsabs.harvard.edu/abs/2019MNRAS.482.4713Z} {482, 4713}

\bibitem[\protect\citeauthoryear{{de Vita}, {Trenti}  \& {MacLeod}}{{de Vita}
  et~al.}{2018}]{deVita18}
{de Vita} R.,  {Trenti} M.,   {MacLeod} M.,  2018, \mn@doi [\mnras]
  {10.1093/mnras/stx3261}, \href
  {https://ui.adsabs.harvard.edu/abs/2018MNRAS.475.1574D} {475, 1574}

\bibitem[\protect\citeauthoryear{{van Leeuwen}, {Le Poole}, {Reijns}, {Freeman}
   \& {de Zeeuw}}{{van Leeuwen} et~al.}{2000}]{vanleeuwen00}
{van Leeuwen} F.,  {Le Poole} R.~S.,  {Reijns} R.~A.,  {Freeman} K.~C.,   {de
  Zeeuw} P.~T.,  2000, \aap, \href
  {https://ui.adsabs.harvard.edu/abs/2000A&A...360..472V} {360, 472}

\bibitem[\protect\citeauthoryear{{van de Ven}, {van den Bosch}, {Verolme}  \&
  {de Zeeuw}}{{van de Ven} et~al.}{2006}]{vandeven06}
{van de Ven} G.,  {van den Bosch} R.~C.~E.,  {Verolme} E.~K.,   {de Zeeuw}
  P.~T.,  2006, \mn@doi [\aap] {10.1051/0004-6361:20053061}, \href
  {https://ui.adsabs.harvard.edu/abs/2006A&A...445..513V} {445, 513}

\bibitem[\protect\citeauthoryear{{van der Marel} \& {Anderson}}{{van der Marel}
  \& {Anderson}}{2010}]{vdMA10}
{van der Marel} R.~P.,  {Anderson} J.,  2010, \mn@doi [\apj]
  {10.1088/0004-637X/710/2/1063}, \href
  {https://ui.adsabs.harvard.edu/abs/2010ApJ...710.1063V} {710, 1063}

\makeatother
\end{thebibliography}

%\begin{thebibliography}{99}
%\end{thebibliography}

%%%%%%%%%%%%%%%%%%%%%%%%%%%%%%%%%%%%%%%%%%%%%%%%%%

%%%%%%%%%%%%%%%%% APPENDICES %%%%%%%%%%%%%%%%%%%%%

\appendix
\renewcommand\thefigure{\thesection A.\arabic{figure}} 
\label{app}
 \begin{figure*}
    \centering
    \includegraphics[trim={1cm 2cm 2cm 1cm},clip,width=\linewidth]{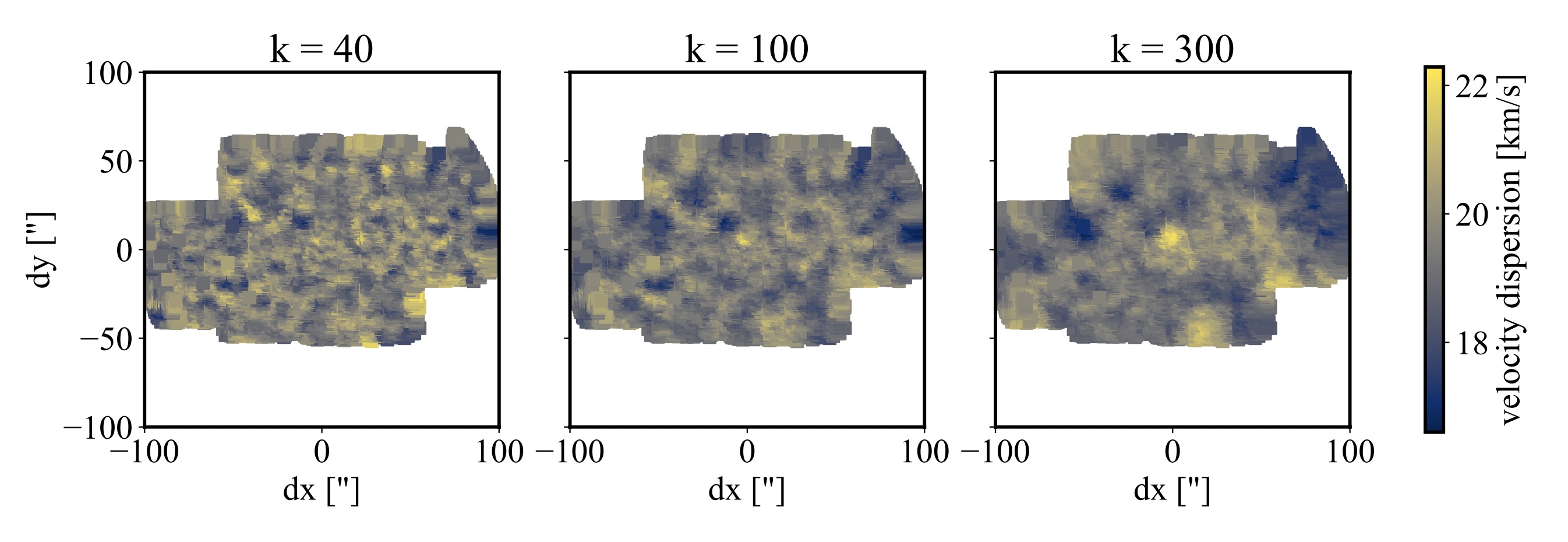}

    \caption{Velocity dispersion maps of $\omega$~Centauri. The maps describe the velocity dispersion determined using the K-nearest neighbors method. From left to right the k--value increases from 40 to 300 in the KNN analysis. A central peak is visible for the k--value of 300. This peak coincides with the center from \citetalias{N10} since their center was derived based on the peak of the velocity dispersion. 
 }
    \label{fig:disp}
\end{figure*}

\section{Appendix}

%%%%%%%%%%%%%%%%%%%%%%%%%%%%%%%%%%%%%%%%%%%%%%%%%%

% Don't change these lines
\bsp	% typesetting comment
\label{lastpage}
\end{document}